\documentclass{article}

\usepackage{amsmath}  
\usepackage[english]{babel}
\usepackage[utf8]{inputenc}
\usepackage{amssymb}
\usepackage{graphicx}
\usepackage{authblk}
\usepackage{caption}
\usepackage{multirow}
\usepackage{amscd}
\usepackage{geometry}
\usepackage{a4wide}
\usepackage[usenames]{color}
\usepackage{etaremune} 
\usepackage{hyperref}
\usepackage{amsfonts}
\usepackage{ifthen}
\usepackage{color}
\usepackage{amscd}
\usepackage{latexsym}
\usepackage{subfigure}

\newcommand{\bd}{\begin{definition}}
	\newcommand{\ed}{\end{definition}}
\newcommand{\bt}{\begin{theorem}}
	\newcommand{\et}{\end{theorem}}
\newcommand{\bi}{\begin{itemize}}
	\newcommand{\ei}{\end{itemize}}
\newcommand{\ben}{\begin{enumerate}}
	\newcommand{\een}{\end{enumerate}}
\newcommand{\beq}{\begin{equation}}
\newcommand{\eeq}{\end{equation}}

\newcommand{\R}{\mbox{$ \mathbb{R}  $}}

\newtheorem{definition}{Def.}[section]
\newtheorem{theorem}{Theorem}[section]
\newtheorem{proposition}{Proposition}[section]

\def \proof{\noindent{\it Proof. \;}  \ignorespaces}
\def \qed{ \hfill $\Box$ \\}

\def \proof{\noindent{\it Proof}.  \ignorespaces}
\def \qed{ \hfill $\Box$ \\}

\title{The relativity of color perception}

\author[1]{Michel Berthier\thanks{michel.berthier@univ-lr.fr}}
\author[2]{Valérie Garcin\thanks{valerie.garcin@math.u-bordeaux.fr}}
\author[3]{Nicoletta Prencipe\thanks{nicoletta.prencipe@math.u-bordeaux.fr}}
\author[4]{Edoardo Provenzi\thanks{edoardo.provenzi@math.u-bordeaux.fr}}
\affil[1]{Laboratoire MIA, Pôle Sciences et Technologie, Université de La Rochelle, Avenue Michel Crépeau, 17042 La Rochelle Cedex 1}
\affil[2,3,4]{Université de Bordeaux, CNRS, Bordeaux INP, IMB, UMR 5251\\ F-33400, 351 Cours de la Libération, Talence, France}

\date{}

\begin{document}
	
\maketitle


\begin{abstract}  
	Physical colors, i.e. reflected or emitted lights entering the eyes from a visual environment, are converted into perceived colors sensed by humans by neurophysiological mechanisms. These processes involve both three types of photoreceptors, the LMS cones, and spectrally opponent and non-opponent interactions resulting from the activity rates of ganglion and lateral geniculate nucleus cells. Thus, color perception is a phenomenon inherently linked to an experimental environment (the visual scene) and an observing apparatus (the human visual system). This is clearly reminiscent of the conceptual foundation of both relativity and quantum mechanics, where the link is between a physical system and the measuring instruments. The relationship between color perception and relativity was explicitly examined for the first time by the physicist H. Yilmaz in 1962 from an experimental point of view. The main purpose of this contribution is to present a rigorous mathematical model that, by taking into account both trichromacy  and color opponency, permits to explain on a purely theoretical basis the relativistic color perception phenomena argued by Yilmaz. Instead of relying directly on relativistic considerations, we base our theory on a quantum interpretation of color perception together with just one assumption, called trichromacy axiom, that summarizes well-established properties of trichromatic color vision within the framework of Jordan algebras. We show how this approach allows us to reconcile trichromacy with Hering's opponency and also to derive the relativistic properties of perceived colors without any additional mathematical or experimental assumption. In doing so, we also introduce several novel and  mathematically rigorous definitions of chromatic attributes and discuss their counterparts in classical colorimetry. Finally, we underline the important role played by the Hilbert metric in our framework and its compatibility with known experimental data.
\end{abstract}

	\section {Introduction} 
In the paper \cite{Yilmaz:62}, H. Yilmaz, inspired by the mathematical physics of the special theory of relativity, explained how to derive colorimetric analogues of Lorentz transformations by exploiting the results of three color perception experiments. However, Yilmaz experiments have remained quite controversial because neither a precise apparatus description nor quantitative data are available. 

In this paper, we overcome this problem by providing a completely theoretical proof of the experimental outcomes claimed by Yilmaz in the setting of the quantum-like framework for color perception introduced and exploited in the papers \cite{BerthierProvenzi:19,Berthier:2020,Berthier:2021,Berthier:2021JofImaging}. As we will explain in section \ref{sec:defquantperceptual}, this framework relies mathematically on just one assumption, that we call `trichromacy axiom', which is meant to sum up in a minimalist mathematical language what is known about the trichromatic aspects of color perception. The trichromacy axiom can be seen as an extended version of the well-established perceptual observations performed by Newton, Maxwell, Grassmann and von Helmholtz, elegantly summarized by Schrödinger in \cite{Schroedinger:20}, and of the theoretical results about homogeneity obtained by Resnikoff in \cite{Resnikoff:74}. 

The mathematical setting of the trichromacy axiom relies on the properties of simple, non-associative, formally real Jordan algebras  $\cal A$ of dimension 3. If $\cal A$ is interpreted as the observable algebra of a quantum theory, then, thanks to the duality state-observable and the density matrix formalism, it is possible to show that the sole trichromacy axiom is sufficient to bring out the chromatic opponent mechanisms advocated by Hering \cite{Hering:1878}.

Several motivations in favor of a quantum theory of color perception, also suggested by the great theoretical physicist A. Ashtekar and his collaborators in \cite{Ashtekar:99}, can be found in \cite{Berthier:2021JofImaging}, where even uncertainty relations for chromatic opponency have been predicted. More motivations will be provided in the discussion section  \ref{sec:discussion}. Here we limit ourselves to quote the illuminant words of B. Russell \cite{Russell:2001} and P.A.M. Dirac \cite{Dirac:82}.

Russel's: ``\textit{When, in ordinary life, we speak of the colour of the table, we only mean the sort of colour which it will seem to have to a normal spectator from an ordinary point of view under usual conditions of light. But the other colours which appear under other conditions have just as good a right to be considered real; and therefore, to avoid favoritism, we are compelled to deny that, in itself, the table has any one particular colour}". 

Dirac's: \textit{``Science is concerned only with observable things and that we can observe an object only by letting it interact with some outside influence. An act of observation is thus necessary accompanied by some disturbance of the object observed"}, and also: \textit{``Questions about what decides the photon’s direction of polarization when it does go through cannot be investigated by experiment and should be regarded as outside the domain of science"}. 

These points of view are clearly much more reminiscent of the way one addresses the problem of measurement in quantum mechanics,  rather than in classical mechanics. 

We consider quite remarkable the fact that the mathematical setting of the trichromacy axiom, together with the quantum interpretation, not only exhibits the intrinsic relationship between opponency and trichromacy, but also, as it will be proven in this paper, its relativistic nature, thus providing a coherent framework for a relativistic quantum theory of color perception in the very simple observational conditions that will be specified in the following sections (and which are common to the great majority of color theories). A complete theory of color perception for observers embedded in natural scenes is still far from being achieved.

The outline of the paper is as follows. Section \ref{sec:Yilmaz} is devoted to the description of Yilmaz contribution \cite{Yilmaz:62}. Our aim in this section is to follow Yilmaz as close as possible. Nevertheless, we adapt the presentation and the argumentation in order to emphasize what information is really taken into account and how it can be used from the mathematical viewpoint. This motivates the introduction of a new nomenclature in section \ref{sec:nomenclaturespecialcolor} with, in particular, the precise definition of an \textit{observer adapted to an illuminant} in the context considered by Yilmaz. In section \ref{sec:Yilmazissues}, we discuss why the outcome of Yilmaz experiments is considered controversial. 

In section \ref{sec:defquantperceptual} the quantum framework for color perception and the associated nomenclature is recalled. Using this setting, we prove in section \ref{sec:Addlaw} the first two experimental outcomes claimed by Yilmaz by using only the trichromacy axiom. The main source of inspiration that guided us during this task is represented by the remarkable Mermin's paper \cite{Mermin:84}, in which it is shown that the core aspect of special relativity is better understood if one concentrates not on Lorentz transformations but on the  Einstein-Poincaré addition law of velocity vectors. This leads us directly to the definition, in section \ref{sec:colorimetricdefs}, of the concept of perceptual chromatic vector, alongside with a whole new additional set of definitions regarding perceptual chromatic attributes. 

The purely theoretical proof that perceptual chromatic vectors satisfy the Einstein-Poincaré addition law, performed in section \ref{sec:EPaddlaw}, allows us to provide a simple mathematical explanation of the outcomes of Yilmaz experiments in \ref{sec:Yilmazproof} and also to point out the relevance of the Hilbert hyperbolic metric on the space of perceptual chromatic vectors: in this context, the Hilbert distance expresses a chromatic constancy property with respect to observer changes, in a sense that will be precisely formalized in section \ref{subsec:metric}.  Our theoretical results are shown to be coherent with existing experimental data in section \ref{subsec:expvalid}.

In section \ref{sec:boostaberration}, we explain how to theoretically recover the outcome of the third Yilmaz experiment, which is crucial in his approach since it avoids resorting to a hypothetical perceptual invariant Minkowski-like quadratic form. Yilmaz already noticed that the corresponding chromatic effect can be considered as an analogue of relativistic aberration. To recover this effect, we essentially show that pure perceptual chromatic states generate one-parameter subgroups of Lorentz boost maps. This is a key result that links the quantum dynamics of chromatic opponency with the relativity of color perception that is  further analyzed in the discussion section \ref{sec:discussion}, in which we provide more justifications for the pertinence of a quantum theory of color perception, with a particular emphasis on its probabilistic interpretation.

Before starting with our original contributions, we discuss in subsection \ref{sec:star} the use of hyperbolic metrics in the color literature. This will allow us to better stress that, differently to all the other works that we have consulted, in our model the Hilbert metric emerges naturally from the mathematical formalism and it is not superimposed to fit experimental data or perceptual effects.

\subsection{State of the art on the use of hyperbolic metrics in color science}\label{sec:star}

Color perception in humans is originated by light spectra, which are  superpositions of  finite-energy electromagnetic waves with wavelengths in the visual spectrum, usually taken to be the interval $\Lambda=[380,780]$, measured in nanometers, and their mathematical representation is given by either positive-valued elements of $L^2(\Lambda)$, or, in the case of monochromatic lights, by singular Dirac-like distributions.

The fact that light spectra, also called color \textit{stimuli}, and color \textit{sensations} are two very distinct concepts has been known since the nineteenth century. Maxwell's experimental observation showed that only three `primaries' are necessary and sufficient to color match any other light spectrum, where three light spectra are called primaries if no linear combination of two of them color matches the remaining one. This fact led first Young and then von Helmholtz to build their celebrated trichromatic theory of color vision.

While the space of physical light spectra is infinite dimensional, that of perceived colors is confined in a three-dimensional space. Nowadays, thanks to  physiological evidences, we know that the biological reason underlying this huge dimensional reduction is that the variability of our photoreceptors is limited to the three LMS cones and that infinitely different light spectra produce the same LMS outputs, thus igniting the same chain of events that leads to a color sensation. This phenomenon is  synthetically referred to as  \textit{metamerism}. Crucially, the post-cones visual chain includes the interlacing of LMS signals, mainly performed by ganglion cells, which gives rise to the achromatic plus color opponent encoding that is sent to the visual cortex and which provides the biological explanation of Hering's theory and also its compatibility with trichromacy.

As a result, the mathematical description of perceived colors is much more involved than that of light spectra and it has been the subject of several proposals. A thorough description of color spaces  is beyond the scope of our paper, here we limit ourselves to compare the different methodologies through which hyperbolic structures are introduced on differently constructed color spaces.

It is possible to highlight  essentially two different ways to mathematically implement the dimensional reduction from the space of light spectra to a color space equipped with a coordinate system: the first, and by far the most widely spread, consists in the CIE (Commission International de l'Éclairage) construction, embedded in a rigorous mathematical framework by Krantz in \cite{Krantz:75}; the second, proposed by Lenz et al. in \cite{Lenz:05,Lenz:07} makes use of a principal component analysis (PCA) performed on a database of light spectra. 

Since the PCA is a well-known technique, the only construction that has to be recalled is that of Krantz or, equivalently, the CIE one. Krantz makes use of Grassmann's laws \cite{Grassmann:1853,Wyszecky:82}, to give a mathematical structure to the space of light spectra and relates it to a cone embedded in a three-dimensional vector space, which he proves to be unique up to a change of basis. Each basis is related to a different way of coding metamerism. 

This is a mathematical explanation of the construction of CIE color spaces  widely used in literature. Let us describe the CIE procedure properly. 
Let $\lambda \in \Lambda$, $C(\lambda)$ be a color stimulus and  $S_i(\lambda)$, $i=$ L,M,S, be the spectral sensitivity function of the LMS cones. The \textit{cone activation coefficients} related to $C$ are  
\begin{equation}\label{eq:coneact}
	\alpha_i(C)=\int_\Lambda C(\lambda) S_i(\lambda) d\lambda, \quad i=\text{L,M,S}
\end{equation} 
and the set of triplets ${(\alpha_i(C))_{i=\text{L,M,S}}}$, as $C$ varies in the space of color stimuli, is called the LMS space. CIE switched the interest away from the LMS space by fixing three primaries $P_k(\lambda)$ and by  defining the \textit{tristimulus values} of $C(\lambda)$ associated to them, denoted with $T_k(C)$, as the three scalar coefficients that permit to combine the primaries $P_k$ in order to color match $C$, i.e. those satisfying the equation $\alpha_i(C)= \sum\limits_{k=1}^3 T_k(C) \int_\Lambda P_k(\lambda) S_i(\lambda) d\lambda$, for all $i=$ L,M,S. 

CIE defined the so-called \textit{color matching functions} (CMF), $T_k:\Lambda \to \R$ as those  satisfying
\begin{equation}\label{eq:tristact}
	T_k(C) = \int_\Lambda C(\lambda) T_k(\lambda) d\lambda, \quad k=1,2,3.
\end{equation}
If we compare eqs. (\ref{eq:coneact}) and (\ref{eq:tristact}) we see that color matching functions are to tristimulus values what cone sensitivity functions are to cone activation values, but, while the latter functions are fixed, color matching functions can vary by selecting different primaries. As we are going to see, the possibility to modify the CMF according to different needs has been exploited in several occasions. It is very important to underline that changing the CMF leads to a change of basis, i.e. a change of coordinate system of the color space. The choice of different bases does not change the nature of the color space, but, as we will see in the following, there are some choices that are perceptually more pertinent than others. In the sequel we are going to list some relevant examples of bases: CIE RGB, CIE XYZ and the basis adopted by Dr\"osler in \cite{Drosler:94}.

In 1931, CIE defined the `standard observer' by fixing the so-called Wright primaries, see e.g. \cite{Wyszecky:82}, or, equivalently, a set of three specific color matching functions denoted with $\bar r, \bar g, \bar b$. The associated tristimulus values are the elements of the famous CIE RGB space, used e.g. in \cite{Koenderink:72}. It is important to stress that this basis is obtained from $3$ physical primaries and it has no perceptual interpretation, in the sense that there is no differentiation between the three coordinates, all of them are of the same kind.

Not pleased with the negative lobe of $\bar r$, CIE modified the primaries and defined other, completely positive, color matching functions\footnote{The color matching function $\bar{y}$ is very similar to the normalized $S_M(\lambda)$.} denoted with $\bar x,\bar y,\bar z$,  giving rise to the equally famous CIE XYZ space, in which Y plays the role of  `luminance', an attribute roughly associated with the intensity of a color stimulus, which can then be seen as an achromatic component. This basis is obtained from the selection of $3$ virtual primaries, and does not permit to describe perceptual features. 
Nevertheless, CIE XYZ, as CIE RGB, is widely used, in particular in \cite{Silberstein:43,vonShelling:56,Judd:70}.

Notice that neither in the RGB nor in the XYZ space, chromatic opposition is considered, while it is in Dr\"osler's color space, defined in \cite{Drosler:94}. His basis is made up by the Gaussian, which minimizes the uncertainty principle, and its two first moments. Unlike the two previous examples, Dr\"osler's choice permits to encode Hering's opponency mechanism in the coordinate system.  Also Yilmaz makes the same choice in \cite{Yilmaz:1962}.

To perceptually identify a color, it seems appropriate to make the distinction between a component called achromatic and  a chromatic one;  the first one is assumed to be mono-dimensional.
This  fact led to the concept of \textit{chromaticity diagram}  firstly introduced by Maxwell in his Cambridge years (1850-1856). 

Nevertheless, given a certain color space the construction of the chromaticity diagram is as arbitrary as the choice of the basis to construct the color space.  Indeed, postulating the existence of an achromatic information, expressed in the coordinates of the color space, is problematic. Thus, it is essentially an operation of dimensional reduction from $3$ to $2$ and there is not a unique way to perform it.\\
A typical example is how the CIE defined the chromaticity coordinates in the XYZ space, by normalization, as $x=X/(X+Y+Z)$, $y=Y/(X+Y+Z)$, $z=Z/(X+Y+Z)$ and defined the color space $xy$Y as the set of all chromaticity coordinates $(x,y)$ together with the luminance Y of all color stimuli. The choice of the plane $(x,y)$ is arbitrary, as the coordinates $(x,z)$, for instance, would have served analogously.

Dr\"osler in \cite{Drosler:94} is the first one to interpret this procedure of normalization in terms of \textit{projective geometry}. Indeed those kind of normalizations can be seen as different choices of affine charts of a projective space of dimension $2$. 

Most of the authors focus on the problem of defining a metric on the chromaticity diagram\footnote{Some of them, e.g. \cite{Koenderink:72,vonShelling:56}, however, define the metric on the whole three-dimensional space.} that they have constructed. While the CIE$xy$ chromaticity diagram described above is tacitly assumed to inherit the Euclidean metric, several hyperbolic proposals have been done in literature to measure distances on chromaticity diagrams.
Two main different approaches can be identified: a first one more conceptually-based on Weber-Fechner's law, a second one more phenomenological and empirical.

Silberstein, in 1938, pursuing the line element method initiated by von Helmholtz in \cite{Helmholtz:1891}, theorizes in \cite{Silberstein:38} that a perceptual metric on chromaticity diagram should not be Euclidean if one assumes Weber-Fechner's law to hold, i.e.
\begin{equation}
	\Delta S = k \frac{\Delta I}{I},
\end{equation}
where $\Delta S$ is the just noticeable difference (JND) in brightness sensation provoked by the modification of light intensity $\Delta I$ w.r.t. a fixed  background intensity $I$, $k$ being a positive real constant.\\
Notice that Weber-Fechner's law says that the line element must be invariant w.r.t. homothetic transformations. Starting from this statement, Dr\"osler in \cite{Drosler:94,Drosler:95} gets the intuition that the space of perceived colors is projective. In particular he states that Weber-Fechner's law in dimension $1$ represents a projective line element that can be generalized to the whole three-dimensional space in \cite{Drosler:95} and to the chromaticity diagram in \cite{Drosler:94}. On this last, because of its projective nature, the metric turns out to be the Klein metric.

Koenderink and his collaborators in \cite{Koenderink:72} implement Weber-Fechner's law in the RGB color coordinates and come up to a Klein-like metric as Dr\"osler. This means that implementing Weber-Fechner's law is equivalent to have a projective model and metric. Some parameters in the metric that they find are set to fit with the Bezold-Br\"uke effect, i.e. the perceptual change in hue when the intensity of a color stimulus is modified, see also chapter 10 of Koenderink's book \cite{Koenderink:10}.

A more phenomenological approach also leads to the idea that Euclidean geometry is not suitable to describe color dissimilarity and to consider a space with non-zero curvature.

The first experimental evidence of this fact has been provided in 1942 by MacAdam: in \cite{MacAdam:42} he shows that the JND contours in the CIE $xy$ chromaticity diagram are not circles, as one would expect from a Euclidean geometry, but are much better approximated by ellipses. This work had an immediate and profound influence on Silberstein, who, in his 1943 paper \cite{Silberstein:43}, defines a perceptual hyperbolic metric, i.e. a perceptual line element, from the MacAdam ellipses.

MacAdam's work also impacted von Schelling: in the 1956  paper \cite{vonShelling:56} he proposes the first, up to our knowledge, explicit  hyperbolic metric, that he considers more relevant to figure out the perception of color differences.

A further evidence in favor of the non-Euclidean nature of a  perceptual color metric  was provided by Judd in 1970 \cite{Judd:70}: an experimental setup to implement von Helmholtz's line element theory showed that the JND of chroma is larger than the JND of hue, i.e. that  humans are more sensitive to changes in hue than in chroma. To describe this phenomenon, Judd coined the term 
`super-importance of hue differences', also known as `hue super-importance'. This work has  inspired Farup and N\"{o}lle and collaborators. Farup, in \cite{Farup:14}, proposes to equip the $a^*b^*$ chromaticity diagram of the CIELab space with the Poincaré metric, showing that this is coherent with both MacAdam's and Judd's results. N\"olle et al., in \cite{Nolle:13} define a new space that takes into account perceptual attributes in the choice of the coordinates and the hue super-importance. Their color space is a three-dimensional manifold embedded in a four-dimensional complex space.
The Euclidean metric of the four-dimensional complex space turns out to be hyperbolic if restricted to the manifold.

Finally, Lenz et al., in \cite{Lenz:05,Lenz:07}, show that the curve described by a point of the chromaticity diagram according to smooth changes of the illuminant almost fits a geodesic of the Poincaré hyperbolic metric.

Ennis and Zaidi also show in \cite{EnnisZaidi:19} that experiments on perceptual barycenters in several CIE spaces imply that their results do not fit with Euclidean geometry and suggest the use of a hyperbolic one.

To resume, in all the works previously mentioned, the authors who invoke hyperbolicity consider  a perceived color as a point of the color space defined under fixed viewing conditions.
As we have seen, the choice of the basis, i.e. of the coordinate system, in the color space is not mathematically well founded. After that, a metric aiming to have a perceptual meaning is introduced on this color space or on the chromaticity diagram.

As we will see in section \ref{sec:nomquant}, our quantum approach is significantly different. We consider a quantum system whose states represent the preparation of the visual scene and where the perceptual information is not contained in the  perceptual color coordinates, but it is obtained from the duality state-observable. The paradigm shift lies in the fact that the measurement procedure is fundamental to model color information. 

As we will detail in the rest of the paper, the existence of the achromatic component, Hering's opponency and the role of hyperbolicity in color perception appear in a natural way from the tricromacy axiom.

\section{Yilmaz relativity of color perception}\label{sec:Yilmaz}

Yilmaz's paper  \cite{Yilmaz:62} is, to the best of our knowledge, the first contribution that investigates the geometry of color perception from the viewpoint of special relativity. The main Yilmaz goal is to obtain {\em colorimetric Lorentz transformations} by interpreting mathematically the outcomes of three basic experiments. Actually, as we will detail in section \ref{sec:Yilmazissues}, these experiments are quite controversial and this fact gives an even stronger motivation to recast Yilmaz in a rigorous mathematical setting where these experiments can be completely bypassed.

\subsection{Yilmaz colorimetric setting}
In order to analyze the results of {\em color matching experiments}, Yilmaz considers a conical color space that, in our notation, can be written as follows:
\begin{equation}\label{eq:CYilmaz}
	\widetilde{\mathcal C}=\{(\alpha,x,y)\in \mathbb R^3, \ \Sigma^2-\Vert {\bf v}\Vert^2\geq 0,\ \alpha\geq 0\}\ ,
\end{equation}
where $\Sigma$ is a non-negative real constant and, when $\alpha > 0$, ${\bf v}=(v_1,v_2)=(x/\alpha,y/\alpha)$, otherwise, if $\alpha=0$, then also $\textbf{v}$ is null. A color $c$ of $\widetilde{\mathcal C}$ can be viewed both as a point of $\mathbb R^3$ with coordinates $(\alpha,x,y)$ and as a couple $(\alpha,{\bf v})$, where $\alpha$ is a positive real number and $ {\bf v}$ is a vector of $\mathbb R^2$ with Euclidean norm given by $v=\Vert {\bf v}\Vert$ less or equal to $\Sigma$. 

In Yilmaz's context, the norm $v=\sqrt{v_1^2+v_2^2}=\sqrt{x^2+y^2}/\alpha$ represents the {\em saturation} of the color $c$ and satisfies $v\le \Sigma$, hence $\Sigma$ is interpreted as the maximal perceivable saturation. Moreover, the angle defined by $\phi=\arctan(y/x)=\arctan(v_2/v_1)$ represents the {\em hue} of $c$ and the non-negative real $\alpha$ is associated to its {\em lightness}.  The definitions of hue, saturation and lightness of classical colorimetry can be consulted for instance in \cite{Wyszecky:82}.

We use the notation $\widetilde{\mathcal C}$ because, in section \ref{sec:trichromacyaxiom}, we will replace $\widetilde{\mathcal C}$ by a  very similar cone that we will call `trichromacy cone' and denoted with $\cal C$. As it will be underlined in \ref{sec:trichromacyaxiom},  $\cal C$ emerges naturally from the so-called trichromacy axiom and it is intrinsically equipped with a rich algebraic structure. Instead, the cone  $\widetilde{\mathcal C}$ has been proposed by Yilmaz in \cite{Yilmaz:62}  on the basis of not fully convincing mathematical arguments, related both to Fourier analysis and to the will of taking into account color opponency, and it is not naturally endowed with the properties of the trichromacy cone.

The existence of a positive real $\Sigma$, which plays the role of a {\em limiting saturation} `reached by spectral colors',  is one of the fundamental assumptions of Yilmaz. The mathematical formula for saturation given above is the analogue of speed (the magnitude of the velocity vector) in mechanics, thus it seems clear that, from Yilmaz's viewpoint, the limiting saturation $\Sigma$ should be interpreted as an analogue of the speed of light. 

Concisely, the purpose of the three experiments described in \cite{Yilmaz:62} is to show that:

1. color perception is a relativistic phenomenon;

2. the limiting saturation is constant under `illuminant changes';

3. there exists a colorimetric aberration effect which is the analogue of the relativistic one.

It is worth mentioning that Yilmaz does not use any information related to a hypothetical invariant quadratic form. In physics, the introduction of an invariant metric on the Minkowski spacetime is motivated by the experimental evidence about the constancy of the speed of light in vacuum measured by inertial observers, however an analogous result is not, or at least not yet, available in the colorimetric setting. It is arguable that this is the reason why Yilmaz wanted to bypass the introduction of an invariant metric by introducing the results of the third experiment.

Our description and subsequent analysis of Yilmaz's experimental results will be greatly simplified if we set up a novel nomenclature adapted from special relativity.

\subsection{The nomenclature of the relativity of color perception}\label{sec:nomenclaturespecialcolor} 

Without any further specification, we consider a color $c$ as an abstract \textit{coordinate-free} element of the space $\widetilde{\cal C}$. This interpretation is the exact analogue to what we do in Galilean mechanics when we consider the position as an abstract element of the space $\R^3$ without coordinates. For color sensations induced by non-self luminous stimuli, a coordinate system can be introduced in $\widetilde{\mathcal C}$ by considering an illuminant\footnote{More precisely, we should call it a broadband illuminant, i.e. a light source extended over the entire visible spectrum. The reason is that, if we consider a narrow-band illuminant, the so-called Helson-Judd effect enters into play and an observer will experience an \textit{incomplete adaptation}, see e.g. \cite{Fairchild:13}. For the sake of simplicity, we will implicitly consider \textit{an illuminant as broadband without further specifications}.} 
which allows us to identify $c$ and to perform measurements on it. For this reason, here we propose the following definition.

\bd[Illuminant]
An illuminant is \textit{a reference frame $I$ of the space $\widetilde{\mathcal C}$}. 
\ed 

It is well-known, see e.g. \cite{Fairchild:13,Gilchrist:99}, that when a person is embedded for a sufficient time in a visual scene illuminated by $I$, he/she will perceive the surface of an object having non-selective reflectance properties without a color saturation. In this case, we call that person \textit{adapted} to $I$. This consideration naturally leads to the following definition.

\bd[Observer]
We call any couple $o=(c,I)$, such that the color $c\in \widetilde{\mathcal C}$ has zero saturation in the reference frame $I$, an observer adapted to the illuminant $I$, or simply an observer. 
\ed 

Given the analogy between the saturation of a color and the speed of a velocity vector for a mechanical system, we can say that an observer $o=(c,I)$ is characterized by the fact that the color $c$ appears `at rest' in the reference frame $I$. Carrying on the analogy with mechanics, we propose the following final definition.

\bd[Inertial observers]
We call $o_1=(c_1,I_1)$ and $o_2=(c_2,I_2)$ two inertial observers and we denote 
by $(\alpha^1,x^1,y^1)=(\alpha^1,\emph{\textbf{v}}_1)$ and $(\alpha^2,x^2,y^2)=(\alpha^2,\emph{\textbf{v}}_2)$ the coordinates of a generic color in the reference frame $I_1$ and  $I_2$, respectively. 
\ed 

By definition of observer, we have that 
$c_i^i=(\alpha^i,0,0)=(\alpha^i,\textbf{0})$, $i=1,2$. However, given $i,j=1,2$, $i\neq j$, $c_j$ will be described by $o_i$ with a color 
$c^i_j$ represented by  
\begin{equation}\label{eq:relativeobs}
	c^i_j=(\alpha,{\bf v}_{ij}),
\end{equation}
where $\alpha \ge 0$ is a suitable non-negative scalar and ${\bf v}_{ij}=\textbf{v}_{c^i_j}$ verifies $v_{ij}=\|{\bf v}_{ij}\|\le \Sigma$.

\subsection{Yilmaz experiments} 	
Thanks to the nomenclature just introduced, we are now able to give a concise description of Yilmaz experiments, for the original description see \cite{Yilmaz:62, Prencipe:20} or the appendix \ref{sec:appendix}. 

In all three experiments, Yilmaz considers only the case of two inertial observes $o_1=(c_1,I_1)$ and $o_2=(c_2,I_2)$ such that only the first component of the vector $\textbf{v}_{12}$ is non-zero, i.e. ${\bf v}_{c^1_2}={\bf v}_{12}=(v_{12},0)$. 

The first experiment is intrinsic in the system given by the two inertial observes: each one describes the color that is perceived at rest by the other. The outcome claimed by Yilmaz is the following: 
\begin{equation}
	{\bf v}_{c_1^2}=-{\bf v}_{c_2^1} \ .
	\label{first}
\end{equation}
If we assume this result to be correct, then it follows that  \textit{color perception is a relativistic phenomenon} and so an absolute description of the sensation of color is meaningless.

The second and the third experiment involve the two inertial observers defined above in the act of observing a particular color $c\in \tilde{\mathcal C}$ which is described by $o_1$ as having maximal saturation, i.e. $\|\textbf{v}_{c^1}\|=\Sigma$, thanks to the contribution of only one component of the vector $\textbf{v}_{c^1}$, the other being zero. The position of the non-null component distinguishes the second from the third experiment. 

Specifically, the outcome of the second experiment can be summarized as follows:
\begin{equation}
	{\bf v}_{c^1}=(\Sigma,0)\Longrightarrow {\bf v}_{c^2}=(\Sigma,0)\ ,
	\label{second}
\end{equation}
i.e., if $c\in \tilde{\mathcal C}$ is described by $o_1$ has having maximal saturation thanks to the sole contribution of the first component of ${\bf v}_{c^1}$, then the description of $c\in \tilde{\mathcal C}$ performed by $o_2$ is identical. 

Instead, the outcome of the third experiment is the following:
\begin{equation}
	{\bf v}_{c^1}=(0,\Sigma)\Longrightarrow {\bf v}_{c^2}=(-\Sigma\sin\varphi,\Sigma\cos\varphi)\ ,
	\label{third}
\end{equation}
with $\sin\varphi=v_{12}/\Sigma$, so, if $c\in \tilde{\mathcal C}$ is described by $o_1$ has having maximal saturation thanks to the sole contribution of the second component of $\textbf{v}_{c^1}$, then $c$ will be still described by $o_2$ as having maximal saturation since $\|\textbf{v}_{c^2}\|=\left(\Sigma^2(\sin^2 \varphi + \cos^2 \varphi)\right)^{1/2}=\Sigma$, but the hue description will be different. 

As already mentioned, the third experiment is meant to mimic the relativistic aberration effect. We are going to see that this experiment is crucial for the derivation of the colorimetric Lorentz transformations performed by Yilmaz.

Finally, we underline that, if Yilmaz outcomes are assumed to be true, then \textit{colors with limiting saturation are perceived as such by all inertial observers}, which is in clear analogy of the fact that the speed of light is measured as constant by all inertial observers. 

\subsection{Yilmaz derivation of colorimetric Lorentz transformations}\label{sec:YilmazLorentz}
We explain now how to obtain the colorimetric Lorentz transformations from eqs. (\ref{first}), (\ref{second}) and (\ref{third}). In \cite{Yilmaz:62} the coordinate change between $o_1$ and $o_2$ is supposed to be linear. When we take into account the specific choices made by Yilmaz, the coordinate change is given by:
\begin{equation}\label{eq:Origins}
	\left(
	\begin{array}{lll}
		\alpha^2\\
		x^2\\
		y^2
	\end{array}
	\right)=
	\left(\begin{array}{ccc}a_{11} & a_{12} & 0 \\a_{21} & a_{22} & 0 \\0 & 0 & 1\end{array}\right)
	\left(
	\begin{array}{lll}
		\alpha^1\\
		x^1\\
		y^1
	\end{array}
	\right)\ .
\end{equation}
The proof of this fact is quite long and technical and it is not relevant for the purposes of this paper. For this reason, we prefer not to include it here and refer the interested reader to the paper \cite{Prencipe:20}, where all the details can be found. 

\begin{proposition}
	With the notations introduced before, the color coordinate transformation corresponding to an illuminant change is the Lorentz boost along the $x$-direction described by the following equation:
	\begin{equation}\label{eq:colorentz}
		\left(
		\begin{array}{lll}
			\alpha^2\\
			x^2\\
			y^2
		\end{array}
		\right)=
		\left(\begin{array}{ccc}{1\over \sqrt{1-(v_{12}/\Sigma)^2}} & {-v_{12}/\Sigma^2\over \sqrt{1-(v_{12}/\Sigma)^2}} & 0 \\ {-v_{12}\over \sqrt{1-(v_{12}/\Sigma)^2}} & {1\over \sqrt{1-(v_{12}/\Sigma)^2}} & 0 \\0 & 0 & 1\end{array}\right)
		\left(
		\begin{array}{lll}
			\alpha^1\\
			x^1\\
			y^1
		\end{array}
		\right)\ .
	\end{equation}
\end{proposition}

\proof 
Using eq. (\ref{eq:Origins}) and by calculating its inverse, after straightforward computations, we obtain:
\begin{equation}
	{x^2\over\alpha^2}={a_{21}\alpha^1+a_{22} x^1\over a_{11}\alpha^1+a_{12}x^1},\ \ \ 
	{x^1\over\alpha^1}={-a_{21}\alpha^2+a_{11} x^2\over a_{22}\alpha^2-a_{12}x^2}\ .
\end{equation}
As it can be checked in more detail in \cite{Prencipe:20}, the fact that $\textbf{v}_{12}=(v_{12},0)$ and eq. (\ref{first}) are equivalent to:
\begin{equation}
	{a_{21}\over a_{11}}=-v_{12},\ \ \ {-a_{21}\over a_{22}}=v_{12}\ .
\end{equation}
This shows that: $a_{11}=a_{22}$ and $a_{21}=-v_{12}a_{22}$.

The result of the second experiment, eq. (\ref{second}), is equivalent to:
\begin{equation}
	\Sigma={a_{21}+a_{22} \Sigma\over a_{11}+a_{12}\Sigma}\ ,
\end{equation}
which gives: $a_{12}=-(v_{12}a_{22})/\Sigma^2$.

From the third experiment, eq. (\ref{third}), we have:
\begin{equation}
	-\tan\varphi={a_{21}\alpha^1+a_{22}x^1\over y^1}={a_{21}\over \Sigma}\ .
\end{equation}
Since $\sin\varphi=v_{12}/\Sigma$, this implies:
\begin{equation}
	a_{22}={1\over \sqrt{1-(v_{12}/\Sigma)^2}}\ .
\end{equation}
\qed 

It is worth noticing that the derivation of these colorimetric Lorentz transformations proposed by Yilmaz relies only on information given by the ${\bf v}$-component of colors, the only one appearing in eqs. (\ref{first}), (\ref{second}) and (\ref{third}). 
As we will see, in the quantum framework these ${\bf v}$-components correspond to the perceptual chromatic vectors that will be introduced in \ref{sec:colorimetricdefs}.
\subsection{Issues about Yilmaz approach}\label{sec:Yilmazissues}
Without calling into question the great originality of Yilmaz's ideas and the relevance of his results, we deem necessary to underline some issues about the approach that we have reported above.
As mentioned before, the derivation of the colorimetric Lorentz transformations is essentially based on the following assumptions:

-- the space of {\em perceived colors} is the cone $\tilde{\cal C}$, and, in particular, there exists a limiting saturation $\Sigma$;

-- the coordinate changes between inertial observers are linear transformations;

-- the results obtained from the three experiments are considered
as valid.

\noindent	
However, no experimental result, nor apparatus description is available in \cite{Yilmaz:62} and this naturally raises doubts about the actual implementation of the three experiments. Furthermore, while the results of the first two experiments are plausible, the outcome of the third seems completely illusory. In fact, Yilmaz defines the limiting saturation of a color $c=(\alpha,x,y)\in \widetilde{\cal C}$ as a value $\Sigma$ of $\|v\|$ that cannot be perceptually matched with that of any Munsell chip, thus, while this definition permits to \textit{identify} the limiting saturation of a color, it does not allow its \textit{measurement}. As a consequence, eq. (\ref{third}), with its precise analytical form, seems to be an ad-hoc formula used to single out the colorimetric Lorentz transformations (\ref{eq:colorentz}), more than the real outcome of a psycho-physical experiment.

It may be tempting to adopt a more conventional approach to obtain the desired transformations starting, for instance, from the fact that there exists a limiting saturation invariant under observer changes and that the color space is isotropic and homogeneous. However, to go further, it is necessary to introduce an analogue of the Minkowski metric, which Yilmaz circumvents. One may choose to follow the standard path used in special relativity, see e.g. \cite{Landau:71,Lechner:18}, to justify the existence of such a metric. However, while the assumptions that go along with this approach rely on a solid experimental basis for what concerns the Minkowski spacetime, they are far from being either obvious or simple to be tested for the space of perceived colors. 

For this reason, we consider a better solution to follow less conventional, but fully equivalent, approaches to special relativity as, e.g., that of the remarkable Mermin's paper \cite{Mermin:84}, whose main focus is the  Einstein-Poincaré velocity addition law and not Lorentz transformations. As mentioned in the introduction, this alternative approach seems more suitable because the colorimetric effects reported by Yilmaz involve the sole ${\bf v}$-components  (or, equivalently, the sole perceptual chromatic vectors that will be defined in section \ref{sec:colorimetricdefs}). The appropriateness of Mermin's approach is also justified by the fact that, as already declared by the emblematic title `Relativity without light', he deals with relativity without specifically considering the physics of electromagnetic waves, thus providing a more general approach that can also be used in our case.

As we have declared in the introduction, we will show how to recover Yilmaz's results from a purely theoretical point of view, thus avoiding the issues discussed in this subsection, thanks to the quantum framework of color perception that is recalled in the next section.

\section{A quantum framework for color perception}\label{sec:defquantperceptual}


In subsection \ref{sec:nomenclaturespecialcolor} we introduced the nomenclature that allowed us developing the relativistic framework for color perception of section \ref{sec:YilmazLorentz}. In the same way, in subsection \ref{sec:nomquant} we introduce a suitable nomenclature,  inspired by the axiomatic description of physical theories, that will be exploited in \ref{sec:trichromacyaxiom}.

\subsection{The nomenclature of visual perception}\label{sec:nomquant}

The following definitions are adapted from the classical references \cite{Emch:2009,Strocchi:08,Moretti:17}. 

\bd[Nomenclature of physical systems]

The following definitions are conventionally assumed in physics.

\bi 

\item A physical system $\mathcal S$ is described as a setting where one can perform physical measures giving rise to quantitative results in conditions that are as isolated as possible from external influences.

\item Observables in $\mathcal S$ are the objects of measurements. If they form an associative and commutative algebraic structure, then the physical theory is called classical.

\item States of $\mathcal S$ are associated with the ways $\cal S$ is prepared for the measurement of its observables.

\item The expectation value of an observable in a given state of $\mathcal S$ is the average result of multiple measures of the observable conducted in the physical system $\mathcal S$ prepared in the same state.

\ei 

\ed 

Regarding this last definition, we notice that this is the standard experimental way of associating a value to an observable both in classical and in quantum physics for two different reasons: in the former we assume that nature is deterministic and observables have precise values, however, we need to introduce the concept of expectation value because all measurements are affected by errors; in the latter we assume that nature is intrinsically probabilistic and the expectation value is needed to associate to every observable the probability that it will take a given value from a set of admissible outcomes.

Observables characterize a state through their measurements and, vice-versa, the preparation of a particular state characterizes the experimental outcomes that will be obtained. It is common to resume this consideration as the duality state-observable. In the standard Hilbert space formulation of quantum theories, observables are Hermitian operators on a Hilbert space, thus they form an associative but non-commutative algebra and the duality observable-state is mathematically formalized by the Riesz-Markov-Kakutani representation theorem \cite{Rudin:06}. However, as we will see next, in the alternative quantum description proposed by Jordan in \cite{Jordan:32}, observables are elements of a commutative but non-associative Jordan algebra and the duality observable-state in this case is encoded in the self-duality property of the positive cone of this algebra. 

When we deal with a visual perceptual system, as an illuminated piece of paper, or a light stimulus in a vision box, the definitions above remain valid, with two major differences: first, the instruments used to measure the observables are not physical devices, but the sensory system of a human being; second, the results may vary from person to person, thus the average procedure needed to experimentally define the expectation value of an observable in a given state is, in general, observer-dependent. The response of an \textit{ideal standard observer} can be obtained through a further statistical average on the observer-dependent expectation values of an observable in a given state.

If we specialize this idea to the case of color perception, we may give the following colorimetric definitions.

\bd[Nomenclature of visual systems]

The following definitions will be assumed in the paper.

\bi 

\item A perceptual chromatic state is represented by the preparation of a visual scene for psycho-visual  experiments in controlled and reproducible conditions.

\item A perceptual color is the perceptual observable identified with a psycho-visual measurement performed in a given perceptual chromatic state.

\item A perceived color is the expectation value assumed by a perceptual color after psycho-visual measurements.

\ei 

\ed 

We underline that the definition of a perceptual color as an observable associated to a psycho-visual measurement in a given perceptual chromatic state is very different than the physical meaning of the term `color stimulus', i.e. the spectral distribution of a light signal across the visual interval. In fact, such a color stimulus, presented to an observer in different conditions, e.g. isolated or in context, can be sensed as dramatically different perceived colors. Thus, from a perceptual viewpoint, it is very ill-posed to identify a perceptual color with a color stimulus, as also mentioned in \cite{Wyszecky:82}, a classical reference for colorimetry.

\subsection{Trichromacy and quantum color opponency}\label{sec:trichromacyaxiom}
In this section we present the quantum theory of color perception on which the rest of the paper will be based upon. We present just an overview of the results obtained in \cite{Berthier:2020,BerthierProvenzi:19,Provenzi:20}, we refer the reader to these papers for details and explanations, especially in what concerns Jordan algebras.

The path that leads to the so-called \textit{trichromacy axiom} can be succinctly summarized as follows. The classical, and well established, colorimetric experiences of Newton, Grassmann and Helmholtz have been resumed by Schrödinger in a set of axioms that describe the structure of a space designed to represent the set of colors from the trichromatic properties of color perception. These axioms stipulate that this space, denoted $\mathcal C$ from now on, is a regular convex cone of real dimension 3. It is important to note that, although it seems tempting to consider $\mathcal C$ as the space of perceived colors, it is more appropriate to consider it as a space of perceptual observables as defined above, see the discussion \ref{sec:discussion} for more details about this issue. Thus, to avoid confusion, we call $\mathcal C$ {\em the trichromacy cone}. In \cite{Resnikoff:74}, Resnikoff showed that to fully exploit this mathematical structure one needs to add a supplementary axiom, namely the fact that $\mathcal C$ is homogeneous, which means that there
exists a transitive group action on $\mathcal C$, see \cite{Provenzi:20} for an extended analysis of the homogeneity axiom. If we add one more property, the self-duality of $\mathcal C$, then $\mathcal C$ becomes a symmetric cone. According to the  Koecher-Vinberg theorem \cite{Baez:12}, the trichromacy cone $\mathcal C$ can then be seen as the positive cone of a formally real Jordan algebra $\mathcal A$, i.e. as the interior of the set of squares of $\mathcal A$. This motivates the following:
\smallbreak
{\textsc{Trichromacy axiom} \cite{Berthier:2020}: --} {\em The trichromacy cone $\mathcal C$ is the positive cone of a formally real Jordan algebra of real dimension 3.} 
\smallbreak

The idea to recast the study of color perception in the Jordan algebra framework appears already in Resnikoff's contribution \cite{Resnikoff:74}. However, Resnikoff was interested in using this concept to understand brightness and he did not mention a possible quantum interpretation of the classical  Schrödinger axioms. Note that our trichromacy axiom differs from these latter by the fact that we require $\mathcal C$ to be homogeneous and self-dual. Self-duality implies that $\mathcal C$ can also be considered as the state cone associated to the perceptual chromatic state space, as we will explain in the following section, thus emphasizing the observable-state duality on which our approach relies.

The most surprising and intriguing consequences of the trichromacy axiom are provided by the classification theorem of Jordan-von Neumann-Wigner, see for instance \cite{Baez:12}. According to this theorem, the Jordan algebra $\mathcal A$ is isomorphic either to the sum $\mathbb R\oplus\mathbb R\oplus\mathbb R$ or to $\mathcal H(2,\mathbb R)$,  the algebra of $2\times 2$ symmetric matrices with real entries, both equipped with a suitable Jordan product. The positive cone of the sum $\mathbb R\oplus\mathbb R\oplus\mathbb R$ is the product $\mathbb R^+\times\mathbb R^+\times\mathbb R^+$. When endowed with the so-called Helmholtz-Stiles metric:
\begin{equation}
	ds^2=\sum\limits_{i=1}^3 a_i\left(d\xi_i/\xi_i\right)^2\ ,
\end{equation}
$a_i, \xi_i\in \mathbb R^+$, it represents the metric space used in the standard colorimetry, see e.g. \cite{Wyszecky:82}. Since this space has been extensively studied, in the sequel we will concentrate only on the second possibility which, as we will see, contains the quantum structure that we are looking for. 

A first crucial remark is that $\mathcal H(2,\mathbb R)$ is naturally isomorphic, as a Jordan algebra, to the {\em spin factor} $\mathbb R\oplus\mathbb R^2$, whose Jordan product is defined by:
\begin{equation}
	(\alpha_1,{\bf v}_1)\circ(\alpha_2,{\bf v}_2)=(\alpha_1\alpha_2+\langle{\bf v}_1,{\bf v}_2\rangle,\alpha_1{\bf v}_2+\alpha_2{\bf v}_1)\ ,
\end{equation}
where $\alpha_1$ and $\alpha_2$ are reals, ${\bf v}_1$ and ${\bf v}_2$ are vectors of $\mathbb R^2$ and $\langle\ ,\ \rangle$ denotes the Euclidean scalar product on $\mathbb R^2$. An explicit isomorphism of Jordan algebras is given by:
\begin{equation}\label{eq:chiso}
	\chi: (\alpha,{\bf v})\in\mathbb R\oplus\mathbb R^2\longmapsto \left(\begin{array}{cc}\alpha+v_1 & v_2 \\v_2 & \alpha-v_1\end{array}\right)\in\mathcal H(2,\mathbb R)\ ,
\end{equation}
where ${\bf v}=(v_1,v_2).$

The positive cone of the Jordan algebra $\mathcal H(2,\mathbb R)$ is the set of positive semi-definite $2\times 2$ symmetric matrices with real entries. Via the isomorphism $\chi$, it corresponds to the positive cone of the spin factor $\mathbb R\oplus\mathbb R^2$. This latter is given by:
\begin{equation}
	\mathcal C=\{(\alpha,{\bf v})\in\mathbb R\oplus\mathbb R^2,\ \alpha^2-\Vert{\bf v}\Vert^2\geq 0,\ \alpha\geq 0\}\ .
\end{equation}
The cone $\mathcal C$ is an explicit representation of the trichromacy cone that we will use extensively in the rest of the paper.

\subsection{Quantum color opponency}\label{sec:quantcolopp}
$\mathcal H(2,\mathbb R)$ is the algebra of observables of the real analogue of a qubit called a {\em rebit}. The states of this quantum system are characterized by {\em density matrices}, i.e. positive unit trace  elements of $\mathcal H(2,\mathbb R)$. It is easy to verify that a matrix of $\mathcal H(2,\mathbb R)$ written as in eq. (\ref{eq:chiso}) is a density matrix if and only if ${\bf v}=(v_1,v_2)\in \mathcal D=\{v\in \mathbb R^2, \; \|v\|\le 1\}$ and $\alpha=1/2$. Two explicit expressions for density matrices are:
\begin{equation}
	\rho({\bf v})={1\over 2}\left(\begin{array}{cc}1+v_1 & v_2 \\v_2 & 1-v_1\end{array}\right)\ ,
	\label{chromatic}
\end{equation}
and
\begin{equation}\label{eq:Pauli}
	\rho({\bf v})={1\over 2}(Id_2+{\bf v}\cdot\sigma)\equiv {1\over 2}(Id_2+v_1\sigma_1+v_2\sigma_2)\ ,
\end{equation}
where $Id_2$ is the $2\times 2$ identity matrix and $\sigma=(\sigma_1,\sigma_2)$, with:
\begin{equation}
	\sigma_1=\left(\begin{array}{cc}1 & 0 \\0 & -1\end{array}\right)\ \ \ \sigma_2=\left(\begin{array}{cc}0 & 1 \\1 & 0\end{array}\right)\ .
\end{equation}

\begin{figure}[htbp]
	\begin{center}
		\includegraphics[width=8cm]{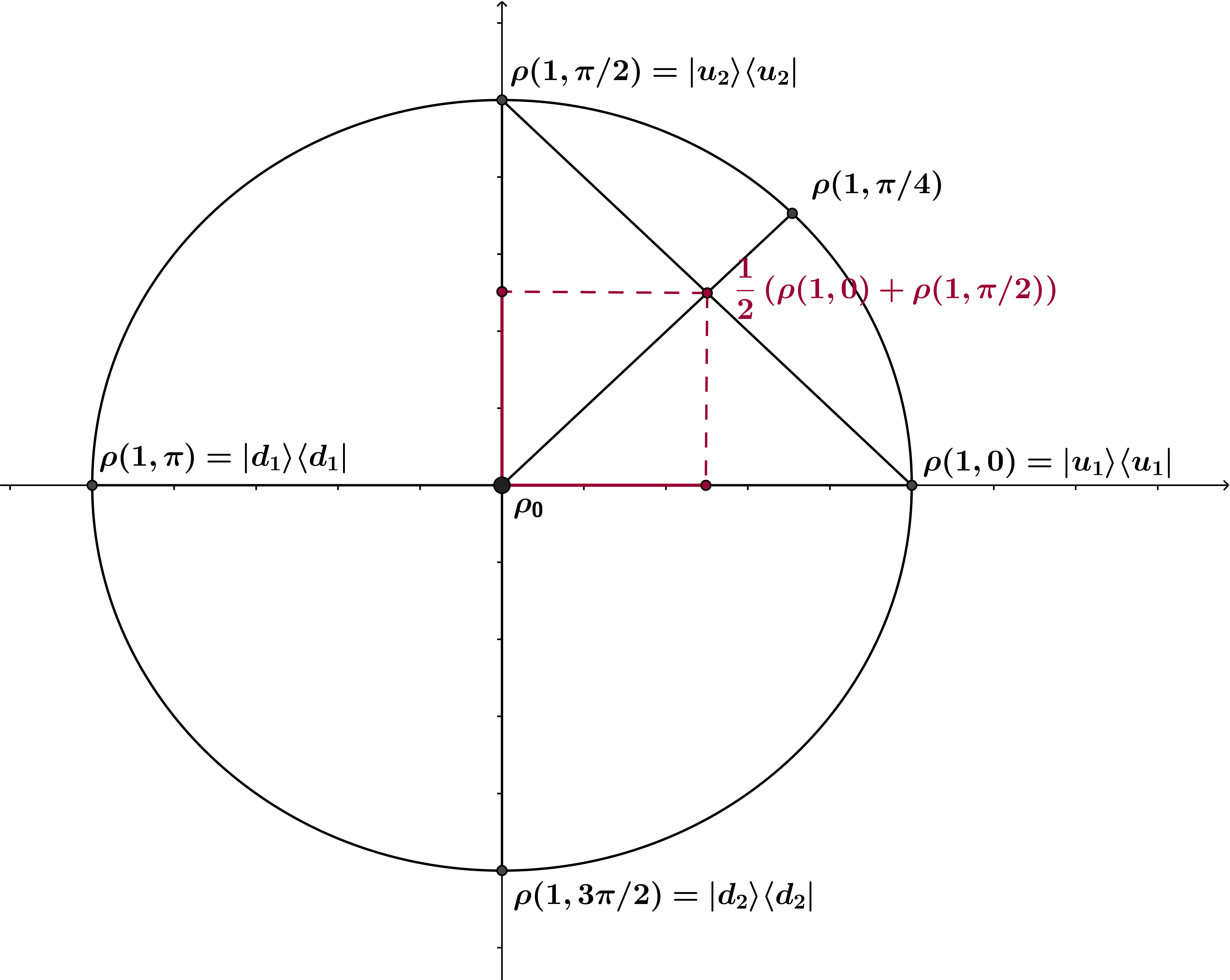}
		\caption{The Bloch disk of the rebit illustrating the opponency mechanism. The density matrix $\rho(1,\pi/4)$ is given by: 
			$\rho(1,\pi/4)=|1,\pi/4\rangle\langle 1,\pi/4|=(|u_1\rangle+|u_2\rangle)(\langle u_1|+\langle u_2|)/(2+\sqrt 2).$
			The mixture $(\rho(1,0)+\rho(1,\pi/2))/2$ is the density matrix: 
			$\rho(\sqrt 2/2,\pi/4)=\rho_0+{1\over 4}(\rho(1,0)-\rho(1,\pi))+{1\over 4}(\rho(1,\pi/2)-\rho(1,3\pi/2)).$}\label{Bloch}
	\end{center}
\end{figure}

The two matrices $\sigma_1$ and $\sigma_2$ are Pauli-like matrices and they are associated to a an opponent mechanism. More precisely, representing $(v_1,v_2)$ in polar coordinates $(r,\theta)$, with $r\in[0,1]$ and $\theta\in[0,2\pi)$, the density matrix $\rho({\bf v})$ can be written in two equivalent forms:
\begin{equation}
	\rho(r,\theta)={1\over 2}\left(\begin{array}{cc}1+r\cos\theta & r\sin\theta \\r\sin\theta & 1-r\cos\theta\end{array}\right)\ ,
\end{equation}
or, by noticing that $\sigma_1=\rho(1,0)-\rho(1,\pi)$ and $\sigma_2=\rho(1,\pi/2)-\rho(1,3\pi/2)$ and using eq. (\ref{eq:Pauli}):
\begin{equation}
	\rho(r,\theta)=\rho_0+{r\cos\theta\over 2}(\rho(1,0)-\rho(1,\pi))+{r\sin\theta\over 2}(\rho(1,\pi/2)-\rho(1,3\pi/2))\ ,
	\label{density}
\end{equation}
where $\rho_0=Id_2/2$ and $\rho(r,\theta)=\rho_0$ if and only if $r=0$. A useful representation of the rebit states is provided by the {\em Bloch disk}, see Figure \ref{Bloch}.

The density matrices parameterized by $r=1$, independently on $\theta\in[0,2\pi)$, i.e. $\rho(1,\theta)$, correspond to {\em pure states}. They are characterized by either the equation: 
\begin{equation}
	\rho(1,\theta)\circ\rho(1,\theta)=\rho(1,\theta)\ ,
\end{equation}
or by: 
\begin{equation}
	-{\rm Trace}(\rho(1,\theta)\log \rho(1,\theta))=0\ ,
\end{equation}
which means that their von Neumann entropy is zero, this latter being defined, for a generic density matrix $\rho$, as $S(\rho)=-\text{Trace}(\rho \log \rho)$. 
The density matrix $\rho_0$ that corresponds to the {\em state of maximal von Neumann entropy} is $\rho_0=Id_2/2$ and it can be written as the mixture:
\begin{equation}
	\rho_0={1\over 4}\rho(1,0)+{1\over 4}\rho(1,\pi)+{1\over 4}\rho(1,\pi/2)+{1\over 4}\rho(1,3\pi/2)\ .
\end{equation}
Eq. (\ref{density}) shows that every density matrix is the sum of the state of maximal entropy with two other components that describe the opponency with respect to the two directions $(\rho(1,0),\rho(1,\pi))$ and $(\rho(1,\pi/2),\rho(1,3\pi/2))$. Given a density matrix $\rho(r,\theta)$, one can evaluate the contribution of the opposition $(\rho(1,0),\rho(1,\pi))$ given by $\sigma_1$ by computing:
\begin{equation}
	\langle\sigma_1\rangle_{\rho(r,\theta)}={\rm Trace}(\rho(r,\theta)\circ\sigma_1)=r\cos\theta\ ,
\end{equation}
and the same for the other direction. It is quite remarkable that the Bloch disk of Figure \ref{Bloch} gives a quantum analogue of the {\em Hering disk} that describes the color opponency mechanism resulting from the activity of certain retinal neurons \cite{Shevell:2017}. The matrix $\sigma_1$ encodes the opposition red/green, while the matrix $\sigma_2$ encodes the opposition yellow/blue. We underline that this quantum justification of the color opponency derives only from the trichromacy axiom when considering the algebra $\mathcal H(2,\mathbb R)$.

\section{Einstein-Poincaré's addition law for chromatic vectors and the formalization of Yilmaz first two experimental results}\label{sec:Addlaw}
In this section we show that the outcomes of the first two experiments quoted by Yilmaz in his model can be rigorously derived from the fact that the so-called perceptual chromatic vectors, that will be introduced in subsection \ref{sec:colorimetricdefs}, satisfy the Einstein-Poincaré addition law. 

As we have done in subsections \ref{sec:nomenclaturespecialcolor}  and \ref{sec:nomquant}, in order to show in the clearest way how to obtain the results stated above, we first need to introduce several notions in subsection \ref{sec:colorimetricdefs}.

\subsection{The nomenclature of quantum perceptual color attributes}\label{sec:colorimetricdefs}
We recall that a {\em perceptual color} $c$ is an element of the trichromacy cone $\cal C$, i.e. explicitly $c=(\alpha,{\bf v})$ with $\alpha^2-\Vert {\bf v}\Vert^2\geq 0$ and $\alpha\geq 0$. 

\bd[Magnitude of a perceptual color]
Let $c=(\alpha,{\bf v})\in \cal C$ be a perceptual color. The positive real $\alpha$ is called the magnitude\footnote{We prefer not to use the term {\em lightness} because of possible confusion. See for instance \cite{Kingdom:2011} for a discussion on the meaning of this word.} of $c$. 
\ed 
Since the cone $\mathcal C$ is self-dual, $c$ can also be considered as an element of the dual cone $\mathcal C^*$.

The case when $c$ has magnitude $\alpha=1/2$ is special, in fact, as previously seen, thanks to the isomorphism defined in eq. (\ref{eq:chiso}), $c$ can naturally be associated to a density matrix representing its state. This justifies the following definition.

\bd[Perceptual color state]
If the perceptual color $c=(\alpha,{\bf v})$ has magnitude $\alpha=1/2$, then $c$ is called a perceptual color state and denoted with $c_s$. Thus, every perceptual color state has the following expression:
\begin{equation}
	c_s:=(1/2,\emph{\textbf{v}}), \; \text{with } \, \| \emph{\textbf{v}}\|\le 1/2.
\end{equation}
\ed 

If we want to associate a perceptual color $c$ with magnitude $\alpha \ge 0$, $\alpha \neq 1/2$,  to a density matrix, we must proceed in two steps: the first consists in dividing $c$ by twice the magnitude, i.e. $c/2\alpha=(1/2,{\bf v}/2\alpha)$, which belongs to $\mathcal D_{1/2}=\{c\in\mathcal C,\ \alpha=1/2\}\cong \{\textbf{u}\in \R^2 \; : \; \|\textbf{u}\|\le 1/2\}$. In this way, the new magnitude is correctly set to $1/2$, coherently with eq. (\ref{chromatic}), but we need a second steps to restore the variability of the vector part inside the unit disk, which is easily accomplished by considering $2\textbf{v}_c \in \mathcal D_1 \cong \{\textbf{u}\in \R^2 \; : \; \|\textbf{u}\|\le 1\}$.

The simple procedure just described leads to the following two definitions.

\bd[Perceptual chromatic vector]
Let $c=(\alpha,{\bf v})\in \mathcal C$, then ${\bf v}_c :={\bf v}/2\alpha \in \mathcal D_{1/2}$ is called the perceptual chromatic vector of  $c$.
\ed 
The reason for the name that we have chosen is that ${\bf v}_c $ carries only information about the chromatic attributes of $c$ and not about its magnitude.

\bd[Perceptual chromatic state]
For every perceptual color $c=(\alpha,{\bf v})\in \mathcal C$, the density matrix $\rho(2{\bf v}_c)$ 
\begin{equation}
	\rho(2{\bf v}_c)={1\over 2}\left(\begin{array}{cc}1+2v_{c,1} & 2v_{c,2} \\2v_{c,2} & 1-2v_{c,1}\end{array}\right)\ .
\end{equation}
is called perceptual chromatic state of $c$.
\ed
The difference between a perceptual \textit{color} state and a perceptual \textit{chromatic} state is represented by the fact that, in the first case, the density matrix associated to a color $c$ with magnitude $1/2$ contains all the information about the state of $c$, magnitude included, which is not the case for a chromatic state, where the magnitude $\alpha$ of $c$ does not play any role.

Two noticeable conditions about perceptual chromatic states can be singled out, as formalized in the following definition.

\bd[Pure and achromatic perceptual states and colors]
Let $c=(\alpha,{\bf v})\in \mathcal C$ be a perceptual color:
\bi 
\item the density matrix $\rho(2{\bf v}_c)$ describes a pure perceptual chromatic state if $\|\textbf{\emph{v}}_c\|=1/2$. If that is the case, then $c$ is called a pure perceptual color;
\item the density matrix $\rho(2{\bf v}_c)$ describes the state of maximal von Neumann entropy if $\textbf{\emph{v}}_c=\textbf{\emph{0}}$. If that is the case, then $c$ is said to be an achromatic perceptual color. 
\ei 
\ed 
Geometrically, pure perceptual colors are in one-to-one correspondence with the points of the boundary of the disk $\mathcal D_{1/2}$, while the center of the disk $\mathcal D_{1/2}$ represents achromatic perceptual colors. 


We now introduce the \textit{chromaticity descriptors}, that we will call purities and quantities. For a closer coherence with Yilmaz model, we will consider only colors $c$ whose perceptual chromatic vectors are of the form ${\bf v}_c=(v_c,0)$ with $-1/2\leq v_c\leq 1/2$. 

\bd[Pure opponent chromatic vectors]
The two chromatic vectors ${\bf v}_+=(1/2,0)$ and ${\bf v}_-=(-1/2,0)$ are called pure opponent chromatic vectors.
\ed 
Given a color $c$, its chromatic vector ${\bf v}_c$ divides the segment connecting $\textbf{v}_-$ and $\textbf{v}_+$ (extremes excluded) in two parts, whose lengths are denoted by $p^-(c)$ and $p^+(c)$, where:
\begin{equation}
	p^-(c) = \frac{1}{2}-v_c={1-2v_c\over 2}\in [0,1], \qquad p^+(c)=v_c-\left(-\frac{1}{2}\right)={1+2v_c\over 2} \in [0,1].
\end{equation}

\bd[$\pm$ purity of a perceptual color]
$p^-(c) $ and $p^+(c) $ will be called the $-$ purity  and the $+$ purity of a perceptual color $c$, respectively.
\ed 
The sum of the $-$ and $+$ purity of $c$ is 1, so $\textbf{v}_c$ can be written as the convex combination of the pure opponent chromatic vectors $\textbf{v}_-$ and $\textbf{v}_+$ with weights given by $p^-$ and $p^+$, respectively, i.e.
\begin{equation}
	\textbf{v}_c=p^-(c)\textbf{v}_- + p^+(c)\textbf{v}_+.
\end{equation} 
The term `purity' is particularly appropriate, not only because it involves the pure opponent chromatic vectors, but also because it is reminiscent of the same term appearing in classical CIE colorimetry. Indeed, also the definition of `excitation purity' $p_e$ of a color $c$ carries the information about its position on a straight line, precisely the one joining the equienergy point $w$ (achromatic color) of the CIE 1931 chromaticity diagram with the so-called dominant wavelength of $c$ (represented by a point belonging to the border of the chromaticity diagram). See \cite{Wyszecky:82} for more details.

\bd[Purity ratio]
Given a perceptual color $c\in \mathcal C$,  such that $\vert v_c\vert \neq 1/2$, the non-negative real number 
\begin{equation}
	r(c)={p^-(c) \over p^+(c)}={1-2v_c\over 1+2v_c} ,
\end{equation}
is called the purity ratio of the color $c$.
\ed 
We have:
\begin{equation}\label{eq:vcpurities}
	v_c={1\over 2}\left({p^+(c)-p^-(c)\over p^+(c)+p^-(c)}\right).
\end{equation}
It is obvious that, given two colors $c$ and $d$, we have:
\begin{equation}
	{\bf v}_{c}={\bf v}_{d}\iff p^+(c)=p^+(d)\iff p^-(c)=p^-(d),
\end{equation}
so, two colors with the same purity may differ only by their magnitude. For this reason, it is useful to define a color attribute analogue to purity but which takes into account also the magnitude information that has been lost after the projection on $\mathcal D_{1/2}$. This is done as follows.

\bd[$\pm$ quantity of a perceptual color] 
Let $c=(\alpha,\textbf{v})$ be a perceptual color. We define the $-$ {\em quantity} $q^-(c)$ and the $+$ {\em quantity} $q^+(c)$ of $c$ by the following two non-negative real numbers:
\begin{equation}\label{eq:facts}
	q^-(c)=2\alpha p^-(c)=\alpha (1-2v_c), \quad 
	q^+(c)=2\alpha p^+(c)=\alpha (1+2v_c).
\end{equation}
\ed 
Of course, perceptual colors with magnitude equal to $1/2$, i.e. perceptual color states, are characterized by the fact that their purities and quantities coincide. 

\subsection{Einstein-Poincar\'e addition law for perceptual chromatic vectors and  Yilmaz first two experiments}\label{sec:EPaddlaw}
Now we discuss our main issue: {\em is there a rigorous way to compare two given colors $c$ and $d$ in $\cal C$}? The answer to this question that seems more natural and coherent with the concepts previously defined is to compare $q^-(c)$ with $q^-(d)$ and $q^+(c)$ with $q^+(d)$, that is to compare their $-$ and $+$ {\em quantities}. For this, we have to introduce the following concept. 

\bd[Quantity ratios] 
Given two perceptual colors $c$ and $d$, such that $\vert v_d\vert \neq 1/2$,  the $\pm$ quantity ratios are defined as:
\begin{equation}\label{eq:qratio}
	s^+(c,d)={q^+(c)\over q^+(d)}\ \ \ {\rm and}\ \ \ s^-(c,d)={q^-(c)\over q^-(d)}\ .
\end{equation}
\ed
If we only know the numerical values of $q^\pm(c)$, $q^\pm(d)$ and not their explicit expressions as in eqs. (\ref{eq:facts}), then using the quantity ratio to compare $c$ and $d$ makes sense only if $d$ is a perceptual color state. In fact, and to take an example, if $c$ and $d$ are two  perceptual colors with the same chromatic vector, the ratio, for instance, $s^+(c,d)$ does not give any information about the description of $c$ relatively to $d$ since we do not know the magnitude of $d$. 

Let us consider two arbitrary perceptual colors $c$ and $d$ whose magnitudes and perceptual chromatic vectors are, respectively, $\alpha_c$ and $\alpha_d$, and ${\bf v}_c$ and ${\bf v}_d$, with $v_c>v_d$. We have:
\begin{equation}
	s^+(c,d)={\alpha_cp^+(c)\over\alpha_dp^+(d)}\ \ \ {\rm and}\ \ \ s^-(c,d)={\alpha_cp^-(c)\over\alpha_dp^-(d)}\ .
\end{equation}
In order to describe $c$ with respect to $d$, we have to perform quantity ratios between $c$ and $d_s$, this latter being the color state whose chromatic vector equals ${\bf v}_d$. We write:
\begin{equation}\label{eq:splus}
	s^+(c,d)={q^+(c)\over 2\alpha_d p^+(d_s)}={q^+(c)\over 2\alpha_d q^+(d_s)}={1\over 2\alpha_d}s^+(c,d_s)\ ,
\end{equation}
and the same with the minus sign. 

Now we arrive to a key definition. 

\bd[Relative perceptual chromatic vector]
Let $c,d \in \cal C$ be two perceptual colors and let $d_s$ be the perceptual color state associated to $d$. Then, the relative perceptual chromatic vector is given by ${\bf v}_c^d=(v_c^d,0)$, where 
\begin{equation}
	v_c^d:={1\over 2}\left({s^+(c,d_s)-s^-(c,d_s)\over s^+(c,d_s)+s^-(c,d_s)}\right)\ .
\end{equation}
\ed
The definition of $\textbf{v}_c^d$ is clearly inspired from eq. (\ref{eq:vcpurities}), but here quantity ratios play the role of purities. We also remark  that the second coordinate of the relative perceptual chromatic vector is 0 because of our choice to consider only perceptual chromatic vectors of the type $\textbf{v}_c=(v_c,0)$, as Yilmaz did. In a future work we will generalize this definition in order to encompass also a non-zero second component. 

\begin{proposition}
	With the notation introduced before, it holds that
	\begin{equation}\label{cdaddition}
		v_c^d={v_c-v_d\over 1-4v_cv_d}\ ,
	\end{equation}
	or, equivalently,
	\begin{equation}\label{addition}
		v_c={v_c^d+v_d\over 1+4v_c^dv_d}.
	\end{equation}
\end{proposition}

\proof 
Thanks to (\ref{eq:splus}) we have $s^{\pm}(c,d_s)=2\alpha_d s^{\pm}(c,d)$, which, using (\ref{eq:qratio}), can be re-written as $s^{\pm}(c,d_s)=2\alpha_d q^{\pm}(c)/q^{\pm}(d)$, so 
\begin{equation}
	v_c^d={1\over 2}\left( {q^+(c)q^-(d)-q^-(c)q^+(d)\over q^+(c)q^-(d)+q^-(c)q^+(d)}\right)\ ,
\end{equation}
and, since the ratio cancels out the proportionality between quantities and purities, we obtain:
\begin{equation}
	v_c^d={1\over 2}\left({p^+(c)p^-(d)-p^-(c)p^+(d)\over p^+(c)p^-(d)+p^-(c)p^+(d)}\right)\ .
\end{equation}
We now notice that:
\begin{equation}
	{v_c-v_d\over 1-4v_cv_d}={ {1\over 2}\left({p^+(c)-p^-(c)\over p^+(c)+p^-(c)}\right) - {1\over 2}\left({p^+(d)-p^-(d)\over p^+(d)+p^-(d)}\right)\over 1- {p^+(c)-p^-(c)\over p^+(c)+p^-(c)}\cdot {p^+(d)-p^-(d)\over p^+(d)+p^-(d)}}\ ,
\end{equation}
straightforward algebraic manipulations lead to
\begin{equation}
	\frac{v_c-v_d}{1-4v_cv_d}={1\over 2}\left({p^+(c)p^-(d)-p^-(c)p^+(d)\over p^+(c)p^-(d)+p^-(c)p^+(d)}\right)=v_c^d,
\end{equation}
and, consequently, to eq. (\ref{addition}).
\qed 

In special relativity, the Einstein-Poincaré addition law between two collinear velocity vectors with speed $u_1$ and $u_2$ can be written as follows: 
\begin{equation}
	u_1\oplus_R u_2 = \frac{u_1+u_2}{1+\frac{u_1u_2}{c^2}},
\end{equation}
where $\oplus_R$ is the symbol used to denote the relativistic sum and $c$ is the speed of light. As we have already remarked in section \ref{sec:Yilmaz}, in Yilmaz's model the analogous of $c$ is the limiting saturation $\Sigma$ that, in the context of perceptual chromatic vectors, is equal to $1/2$. This explains the presence of the factor 4 in eqs. (\ref{cdaddition}) and (\ref{addition}), which are the exact analogue of the Einstein-Poincaré addition law for perceptual chromatic vectors written with our nomenclature. In particular, eq. (\ref{addition})  establishes that, given any two perceptual colors $c$ and $d$, the relativistic sum of $v_d$ with the relative perceptual chromatic vector $v_c^d$ leads to $v_c$.

\subsection{A theoretical proof of the first two outcomes of Yilmaz experiments}\label{sec:Yilmazproof}

Thanks to eqs. (\ref{cdaddition}) and (\ref{addition}), we can prove the first two outcomes of Yilmaz's experiments in a purely theoretical manner. The proof of the first one is extremely simple, in fact, by exchanging $c$ and $d$ in eq. (\ref{cdaddition}) we immediately find that 
\begin{equation}\label{eq:Yilmazfirst}
	v^d_c=-v^c_d,
\end{equation}
which is nothing but an alternative way of writing eq. (\ref{first}), i.e. the first experimental outcome claimed by Yilmaz. 
From eq. (\ref{cdaddition}) it follows that $v_d=(v_c - v_c^d)/(1-4v_c^d v_c)$, but thanks to eq. (\ref{eq:Yilmazfirst}) we can also write 
\begin{equation}\label{additiond}
	v_d={v_d^c + v_c\over 1+4v_d^c v_c}.
\end{equation}
The theoretical proof of the second experimental outcome claimed by Yilmaz, i.e. (\ref{second}), is a bit trickier. First of all, we must recall that the second Yilmaz experiment involves two inertial observers $o_1=(c_1,I_1)$ and $o_2=(c_2,I_2)$ perceiving a maximally saturated color, which gives rise to the two vectors $\textbf{v}_{c^1}=(\Sigma,0)$ and $\textbf{v}_{c^2}=(\Sigma,0)$, together with the vector $\textbf{v}_{c^1_2}$, which encodes how $o_1$ describes the color $c_2$. Instead, in this section, we deal with two perceptual colors $c,d\in \cal C$, which are associated to the perceptual chromatic vectors $\textbf{v}_c, \textbf{v}_d \in \mathcal D_{1/2}$, respectively, together with the relative perceptual chromatic vector $\textbf{v}_c^d\in \mathcal D_{1/2}$. Thus, if we want to find a correlation, we must first operate suitable identifications among the three vectors appearing in the two situations. 

Naively, we would be tempted to identify $\textbf{v}_{c^1}$ with $\textbf{v}_c$, $\textbf{v}_{c^2}$ with $\textbf{v}_d$ and $\textbf{v}_{c^1_2}$ with $\textbf{v}_c^d$, however this would not lead to the correct interpretation of the outcome of the second Yilmaz experiment in terms of the results presented in this section. Instead, the correct identifications are the following

\begin{displaymath}
	\begin{cases}
		\textbf{v}_{c^1} \equiv \textbf{v}_c \\
		\textbf{v}_{c^2}  \equiv \textbf{v}_c^d\\
		\textbf{v}_{c^1_2} \equiv  \textbf{v}_{d}
	\end{cases},
\end{displaymath}
in fact, if, for the reasons explained above, we replace $\Sigma$ with $1/2$ and we introduce $v_c=1/2$ in eq. (\ref{cdaddition}), we find that $v_c^d=1/2$ independently of $v_d$. This is the precise way in which the second outcome claimed by Yilmaz must be interpreted within the formalism of perceptual chromatic vectors. 

\section{A theoretically and experimentally coherent distance on the space of perceptual chromatic vectors: the Hilbert metric}
\label{subsec:metric}
In this section we prove that, quite remarkably,  the Einstein-Poincaré additivity law satisfied by perceptual chromatic vectors permits to coherently equip the space of such  vectors with the so-called Hilbert metric. In subsection \ref{subsec:expvalid}, we show that this metric is compatible with the results of well-established psycho-visual experiments.

Let us start by recalling that, given four collinear points $a$, $p$, $q$, and $b$ of $\mathbb R^2$, with $a\neq p$ and $q\neq b$, the \textit{cross ratio} $\left[a,p,q,b\right]$ is defined by \cite{Colbois:07}:
\begin{equation}
	\left[a,p,q,b\right] = \frac{\Vert q-a \Vert}{\Vert p-a \Vert} \cdot \frac{\Vert p-b \Vert}{\Vert q-b \Vert} \ ,
\end{equation}
where $\Vert\cdot\Vert$ denotes the Euclidean norm. Given two points $p$ and $q$ of the closed disk $\mathcal D_{1/2}$ such that the points $(-1/2,0)=a_-$, $p$, $q$, and $(1/2,0)=a_+$ are collinear with the segment $[p,q]$ contained in the segment $[a_-,a_+]$, the $\mathcal D_{1/2}$-Hilbert distance $d_H(p,q)$ is given by \cite{Colbois:07}:
\begin{equation}
	d_H(p,q)={1\over 2}\ln\left[a_-,p,q,a_+\right]\ ,
\end{equation}
where the choice of the points involved in the cross ratio above guarantees that the argument of $\ln$ is strictly positive.

We consider now three chromatic vectors ${\bf v}_c$, ${\bf v}_d$ and ${\bf v}_c^d$ of $\mathcal D_{1/2}$ with ${\bf v}_c=(v_c,0)$, ${\bf v}_d=(v_d,0)$ and ${\bf v}_c^d=(v_c^d,0)$. We have the following result (see for instance \cite{Fock:2015} for related topics).

\begin{proposition} With the notations introduced above, it holds that:
	\begin{equation}
		d_H((0,0),(v_c^d,0))=d_H((v_d,0),(v_c,0))\iff v_c={v_c^d+v_d\over 1+4v_c^dv_d}\ .
		\label{addition2}
	\end{equation}
	\label{prop1}
\end{proposition}

\proof By definition, the equality $d_H((0,0),(v_c^d,0))=d_H((v_c,0),(v_d,0))$ holds if and only if 
$\left[a_-,(0,0),(v_c^d,0),a_+\right]=\left[a_-,(v_d,0),(v_c,0),a_+\right]$.  Equivalently:
\begin{equation}\label{eq:Hilbdist}
	d_H((0,0),(v_c^d,0))=d_H((v_c,0),(v_d,0))\iff {1/2-v_c\over 1/2+v_c}={1/2-v_c^d\over 1/2+v_c^d}\cdot {1/2-v_d\over 1/2+v_d}\ .
\end{equation}
By a straightforward computation, it can be checked that the right-hand side  of \eqref{eq:Hilbdist} is equivalent to that of  \eqref{addition2}.
\qed 

By using the vector notation, \eqref{addition2} can be re-written as follows
\begin{equation}\label{eq:chasles}
	d_H({\bf 0},{\bf v}_c^d)=d_H({\bf v}_d,{\bf v}_c)\iff v_c=\frac{v_c^d+v_d}{1+4v_c^d v_d},
\end{equation}
i.e. the relative perceptual chromatic vector $\textbf{v}_c^d$ appears in the relativistic sum expressed by (\ref{addition}) together with the perceptual chromatic vectors $\textbf{v}_c$ and $\textbf{v}_d$ if and only if the \textit{Hilbert length} $d_H({\bf 0},{\bf v}_c^d)$ of $\textbf{v}_c^d$ is equal to the Hilbert distance between $\textbf{v}_c$ and $\textbf{v}_d$. 

The colorimetric interpretation is the following:  since the relativistic sum (\ref{addition}) has been previously proven to hold true, this result implies that our hypothesis that $\textbf{v}_c^d$ contains information about the perceptual dissimilarity between the colors $c$ and $d$ is verified if and only if we consider the chromatic vectors as elements of the metric space $(\mathcal D_{1/2},d_H)$, thus promoting the  Hilbert distance to a mathematically coherent candidate for a perceptual metric of chromatic attributes.

Remarkably, see e.g. \cite{Beardon:99}, the Hilbert metric on $\mathcal D_{1/2}$ coincides precisely with the {\em Klein hyperbolic metric} defined by:
\begin{equation}
	ds^2_{\mathcal D_{1/2}}={(1/4-v_2^2)dv_1^2+2v_1v_2dv_1dv_2+(1/4-v_1^2)dv_2^2\over (1/4-\Vert v\Vert^2)^2}\ .
\end{equation}
The geodesics with respect to this metric are straight chords of $\mathcal D_{1/2}$.

\begin{figure}[htbp]
	\begin{center}
		\includegraphics[width=7cm]{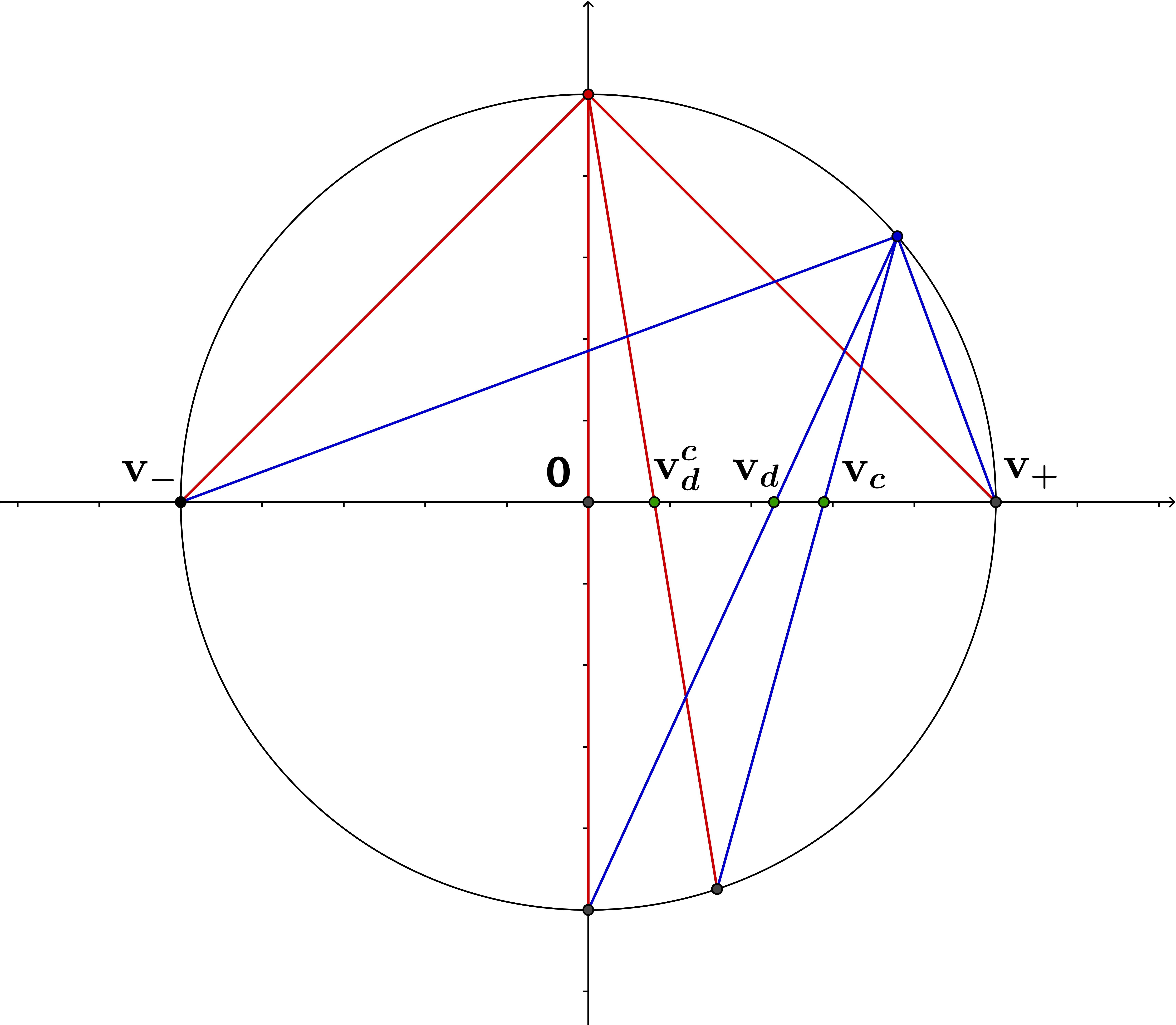}
		\caption{Illustration of the result of Prop. (\ref{prop1}) by Chasles theorem on the cross ratios of cocyclic points. $\textbf{v}_c$, $\textbf{v}_d$ and $\textbf{v}^d_c$ satisfy eq. (\ref{eq:chasles}).}\label{Chasles}
	\end{center}
\end{figure}

A geometric representation of this result is provided by the so-called Chasles theorem on cross ratios of cocyclic points, see Figure \ref{Chasles}, which provides a graphical method to construct the relativistic sum of two vectors in one dimension. 

An alternative interpretation of formula (\ref{eq:chasles}) is possible by recasting it in the context of the inertial observers framework introduced in section \ref{sec:Yilmaz}. To remain coherent with the assumption of section \ref{sec:colorimetricdefs}, we will consider only chromatic vectors of the type $\textbf{v}_c=(v_c,0)$.

Considering again the notation of section \ref{sec:nomenclaturespecialcolor}, let $o_1=(c_1,I_1)$ and $o_2=(c_2,I_2)$ be two inertial observers\footnote{We recall that, for the sake of a simpler phrasing, we implicitly assume that the inertial observer $o_i$ is adapted to the illuminant $I_i$, $i=1,2$, without explicitly specifying it.}, then, by definition, $\textbf{v}_{11}=(0,0)$ and $\textbf{v}_{22}=(0,0)$. However, using the notation introduced in eq. (\ref{eq:relativeobs}), the inertial observer $o_1$ perceives $c_2$ with a non-zero saturation, i.e. $\textbf{v}_{12}=(v_{12},0)$, with $v_{12}\neq 0$, and, thanks to eq. (\ref{eq:Yilmazfirst}), $\textbf{v}_{21}=(-v_{12},0)$.

Furthermore, fixed $F\in \cal C$, let $\textbf{v}_{1F}=(v_{1F},0)$ and $\textbf{v}_{2F}=(v_{2F},0)$ be the chromatic vectors corresponding to the description of $F$ performed by the inertial observers $o_1$ and $o_2$, respectively.

Coherently with the analysis made in section \ref{sec:Yilmazproof}, we perform the following identifications between the chromatic vector components of the colors $c$ and $d$ appearing in formula (\ref{eq:chasles}) and those of $c_1$, $c_2$ and $F$: 
\begin{displaymath}
	\begin{cases}
		\textbf{v}_d \equiv \textbf{v}_{12}\\
		\textbf{v}_c \equiv \textbf{v}_{1F}\\
		\textbf{v}^d_c \equiv \textbf{v}_{2F}
	\end{cases},
\end{displaymath}
then formula (\ref{eq:chasles}) implies the equality
\begin{equation}
	d_H(\textbf{v}_{22},\textbf{v}_{2F})=d_H(\textbf{v}_{12},\textbf{v}_{1F}), 
\end{equation}
notice that the arguments of the Hilbert distance in the left-hand side are relative to the color description performed by $o_2$ and those in the right-hand side are relative to $o_1$.  Since $v_{22}=0$, we can also write
\begin{equation}\label{eq:constancy}
	d_H(\textbf{0},\textbf{v}_{2F})=d_H(\textbf{v}_{12},\textbf{v}_{1F}).
\end{equation}
The interpretation of formula (\ref{eq:constancy}) gives a rigorous meaning to the sentence that we wrote in the introduction about the fact that the Hilbert distance provides a `\textit{chromatic constancy property with respect to observer changes}'. In fact, if we interpret the Hilbert distance as a perceptual metric, eq. (\ref{eq:constancy}) says that the perceptual chromatic difference between  $F$ and an achromatic color sensed by $o_1$ is the same as the one that $o_2$ experiences between $F$ and the chromatic vector $\textbf{v}_{12}$ representing the saturation shift due to the observer change from $o_1$ to $o_2$.

We stress that we have implicitly assumed the  illuminants $I_1$ and $I_2$ to be broadband, so the previous interpretation is valid as long as the quantity $v_d=v_{12}$ is relatively small.

\subsection{Compatibility of the Hilbert metric with psycho-visual experimental data}\label{subsec:expvalid}

\begin{figure}[htbp]
	\begin{center}
		\includegraphics[width=8cm]{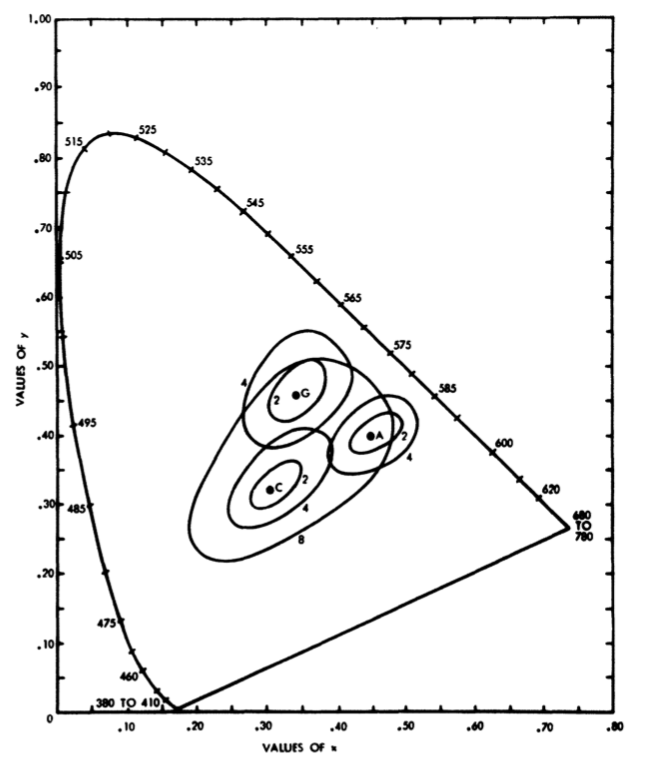}
		\caption{The iso Munsell chroma contours found by  \cite{Crocetti:63} in the $xyY$ diagram.}\label{fig:IsoChroma}
	\end{center}
\end{figure}

Now we address the important issue of the compatibility between the Hilbert metric on $\mathcal D_{1/2}$ and psychovisual measurements. This is not an easy task because of two reasons: firstly, experimental data on color perception are very scarce, secondly, psychovisual measurements are always affected by subjective variations which imply the use of averaging procedures that inevitably reduce the measure accuracy. The only psychovisual results consistent with our framework that we were able to find are those reported in \cite{Burnham:57} and \cite{Crocetti:63}. The authors conducted their tests with the help of the standard CIE illuminants $C$ (near-daylight, $(x_C,y_C)=(0.3125,0.3343)$) and $A$ (tungsten, $(x_A,y_A)=(0.4475, 0.4084)$) and added a third one, denoted with $G$ (greenish, $(x_G,y_G)=(0.3446,0.4672)$). The values $(x,y)$ represent the CIE $xyY$ chromaticity coordinates of $C$, $A$ and $G$, respectively, Fig. \ref{fig:IsoChroma} shows their position in the chromaticity diagram. In what follows, observers adapted to the illuminants $C$, $A$ and $G$, respectively, will be denoted by $o_1=(c,C)$, $o_2=(a,A)$ and $o_3=(g,G)$. A haploscope is used to compare the color perception of one eye always adapted to the illuminant $C$ and the other eye adapted to $C$, $A$ and $G$.

Fig. \ref{fig:IsoChroma} shows, in the $xyY$ diagram, three families of curves obtained by the tests performed in \cite{Crocetti:63}:

-- the first is composed by three contours surrounding $C$ that correspond to color stimuli with fixed Munsell value, different hue but with the same perceived Munsell chroma in $\{2,4,8\}$. By normalizing these data between $0$ and $0.5$ we obtain $\{0.1,0.2,0.4\}$, which are the norms of the chromatic vectors ${\bf v}_{1c}$ of the colors associated to the corresponding stimuli observed by $o_1$;

-- the second and the third are given by two contours surrounding $A$, resp. $G$, that correspond to colors $c$ with varying hues and whose Munsell chroma belong to the set $\{2,4\}$. The chromatic vectors ${\bf v}_{2c}$, resp. ${\bf v}_{3c}$, of these colors observed from $o_2$, resp. $o_3$, have norms belonging to the set $\{0.1,0.2\}$.

As discussed above, the psychovisual data reported in \cite{Burnham:57} and \cite{Crocetti:63} are only averaged, thus, the only kind of information that we have from Fig. \ref{fig:IsoChroma} is, for example, that the $xyY$ coordinates of standard illuminant $A$ are between the curves of chroma 4 and 8 of the observer $o_1$. Thus, the norm of the chromatic vectors is not possible to achieve with accuracy. However, in order to test our mathematical theory, as a first approximation, we perform a linear interpolation from the data appearing in the figure, which gives $\Vert {\bf v}_{1a}\Vert\simeq 6.76/20=0.338$. 

\begin{figure}[h!]
	\centering 
	\subfigure[The illuminants $C$ and $A$ and the colors $F$ and $F'$ in the $xyY$ diagram.]{\includegraphics[width=0.47\linewidth]{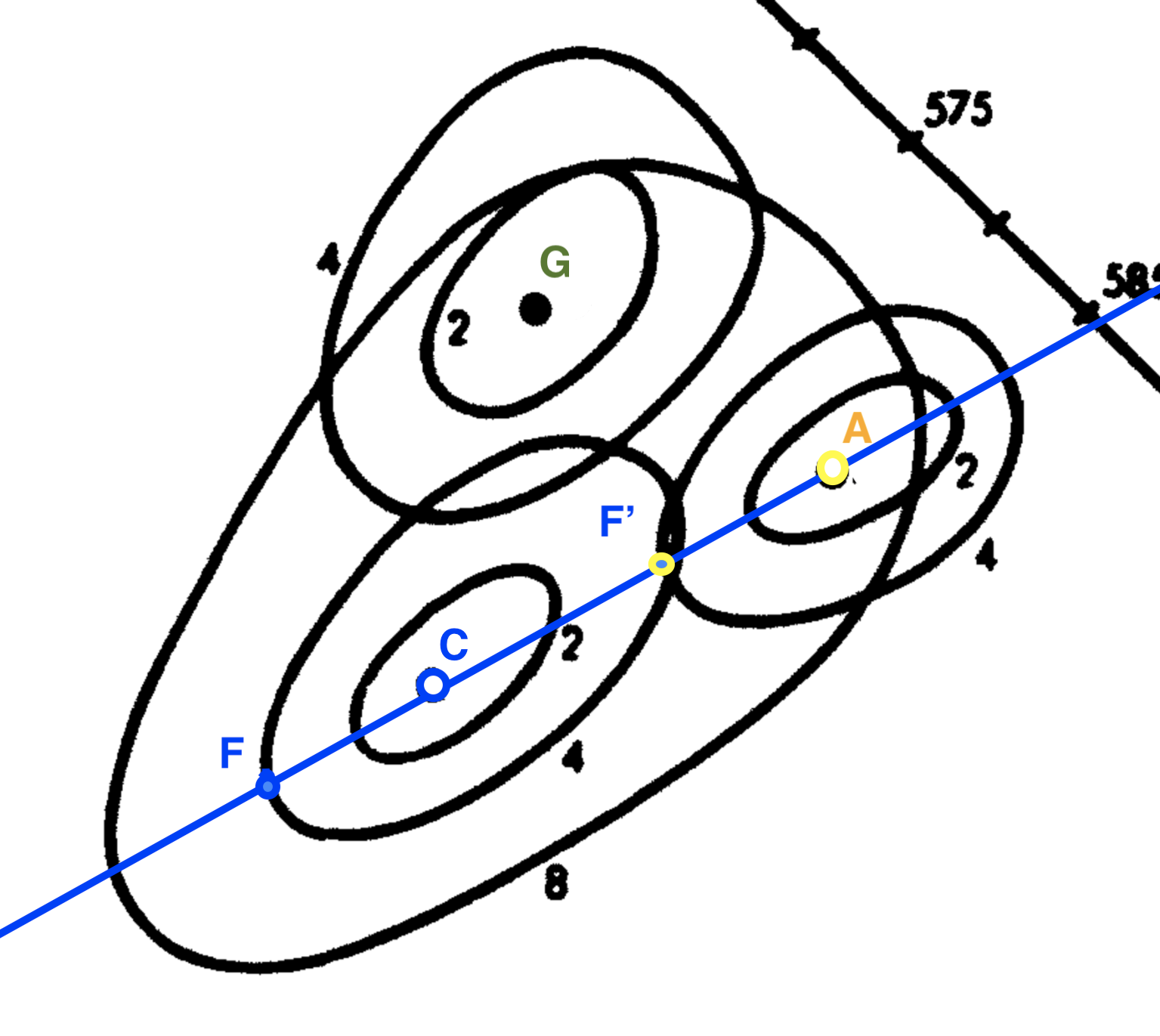}\label{CA_CIE}}
	\hspace{5mm}
	\subfigure[Illustration of the equalities of eq. (\ref{CA_Cross_Chasles}) in the disk $\mathcal{D}_{1/2}$.]
	{\includegraphics[width=0.42\linewidth]{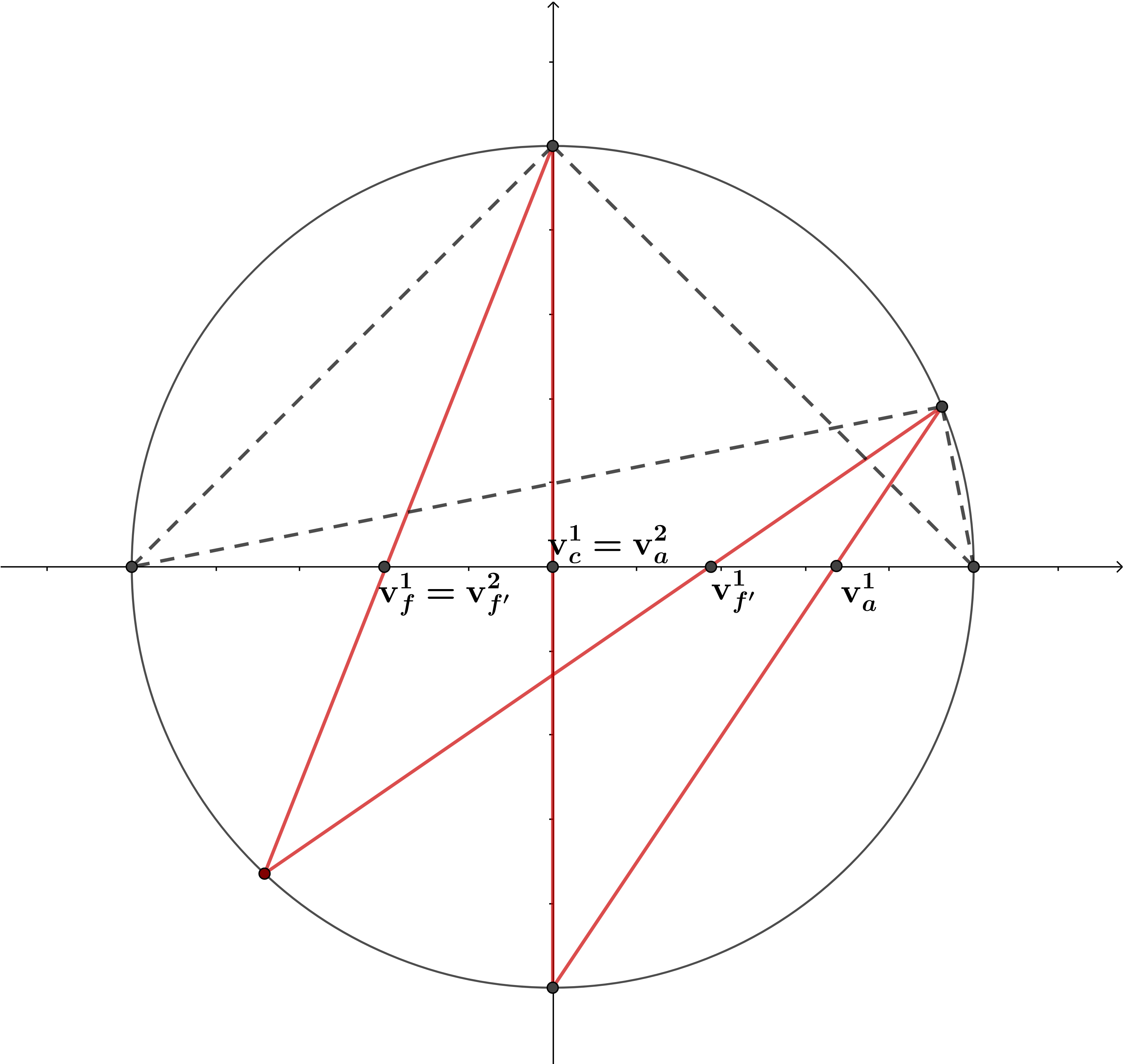}\label{CA_Cross}}
	\caption{Invariance of the Hilbert distance under observer changes: illuminants $C$ and $A$, and colors $F$ and $F'$.}
	\label{HilbertCA}
\end{figure}

In Fig. \ref{CA_CIE}, we denote by $F$ and $F'$ the $xyY$ coordinates of the points in the $xyY$ diagram obtained by the intersection between the line connecting $A$ and $C$ with the iso-chroma contours for $o_1$ and $o_2$, respectively. The color $F$ is perceived by $o_1$ as having a chromatic vector ${\bf v}_{1F}$ with norm $\Vert {\bf v}_{1F}\Vert=0.2$. By construction, we determine $F'$, the color perceived by $o_2$ with chromatic vector ${\bf v}_{2F'}$ such that ${\bf v}_{2F'}={\bf v}_{1F}$. Again, by linear interpolation, the norm of the chromatic vector ${\bf v}_{1F'}$ corresponding to the color $F'$ perceived by $o_1$, is approximated by $\Vert {\bf v}_{1F'}\Vert\simeq 3.76/20=0.188$. Fig. \ref{CA_Cross} shows all the chromatic vectors in the disk $\mathcal D_{1/2}$.

One can easily check, as illustrated by Chasles theorem, that:
\begin{equation}
	d_H (\textbf{v}_{1F},\textbf{v}_{1c}) = d_H (\textbf{v}_{2F'},\textbf{v}_{2a})= d_H (\textbf{v}_{1F'},\textbf{v}_{1a})\ .
	\label{CA_Cross_Chasles}
\end{equation}

\begin{figure}[h!]
	\centering 
	\subfigure[The illuminants $C$ and $A$ and the colors $F$ and $F'$, and $F_2$ and $F_2'$ in the CIE xyY diagram.]{\includegraphics[width=0.47\linewidth]{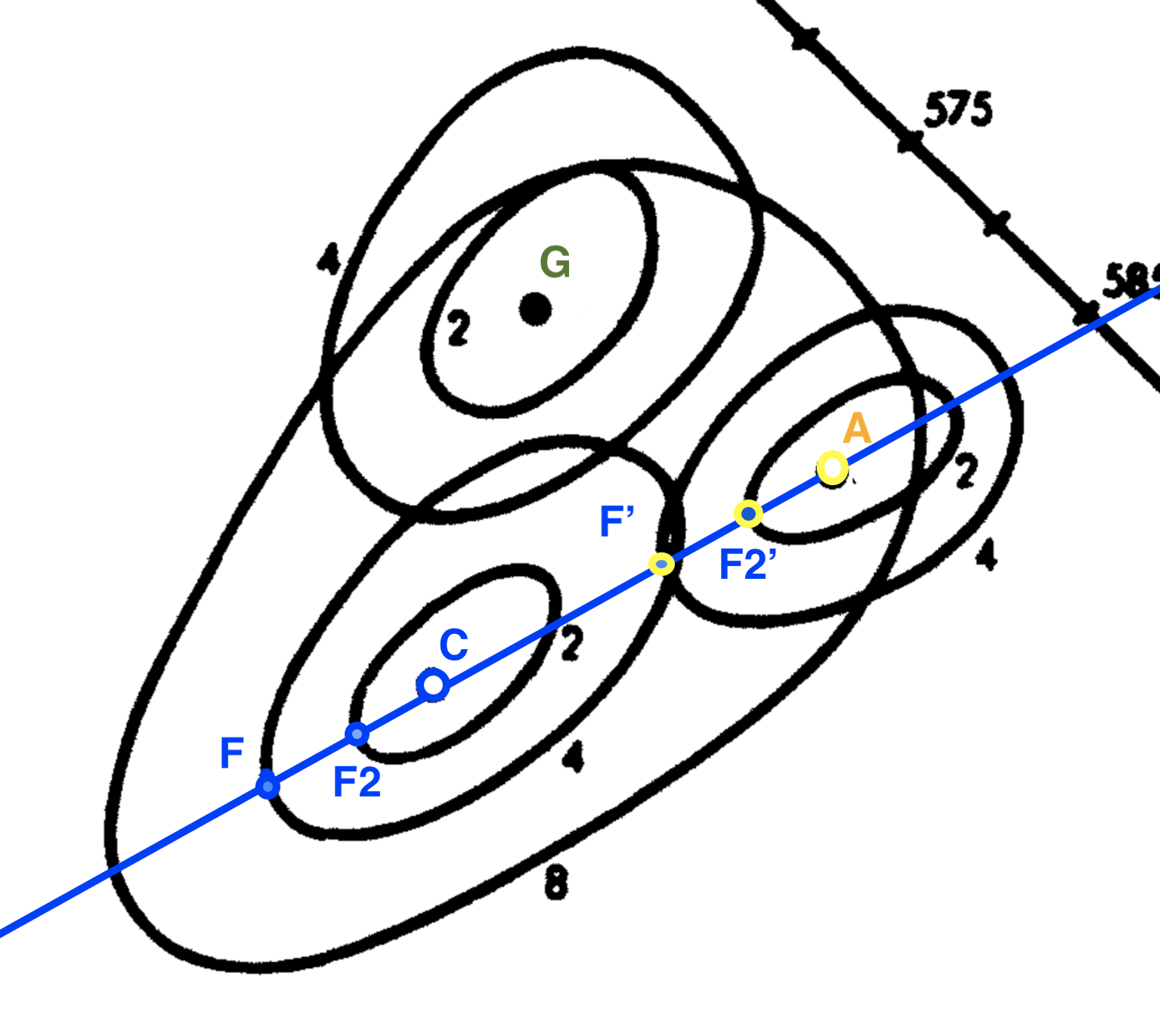}\label{CA_CIE2}}
	\hspace{5mm}
	\subfigure[Illustration of the equalities of Eq. (\ref{CA_Cross_Chasles2}) in the disk $\mathcal{D}_{1/2}$.]
	{\includegraphics[width=0.42\linewidth]{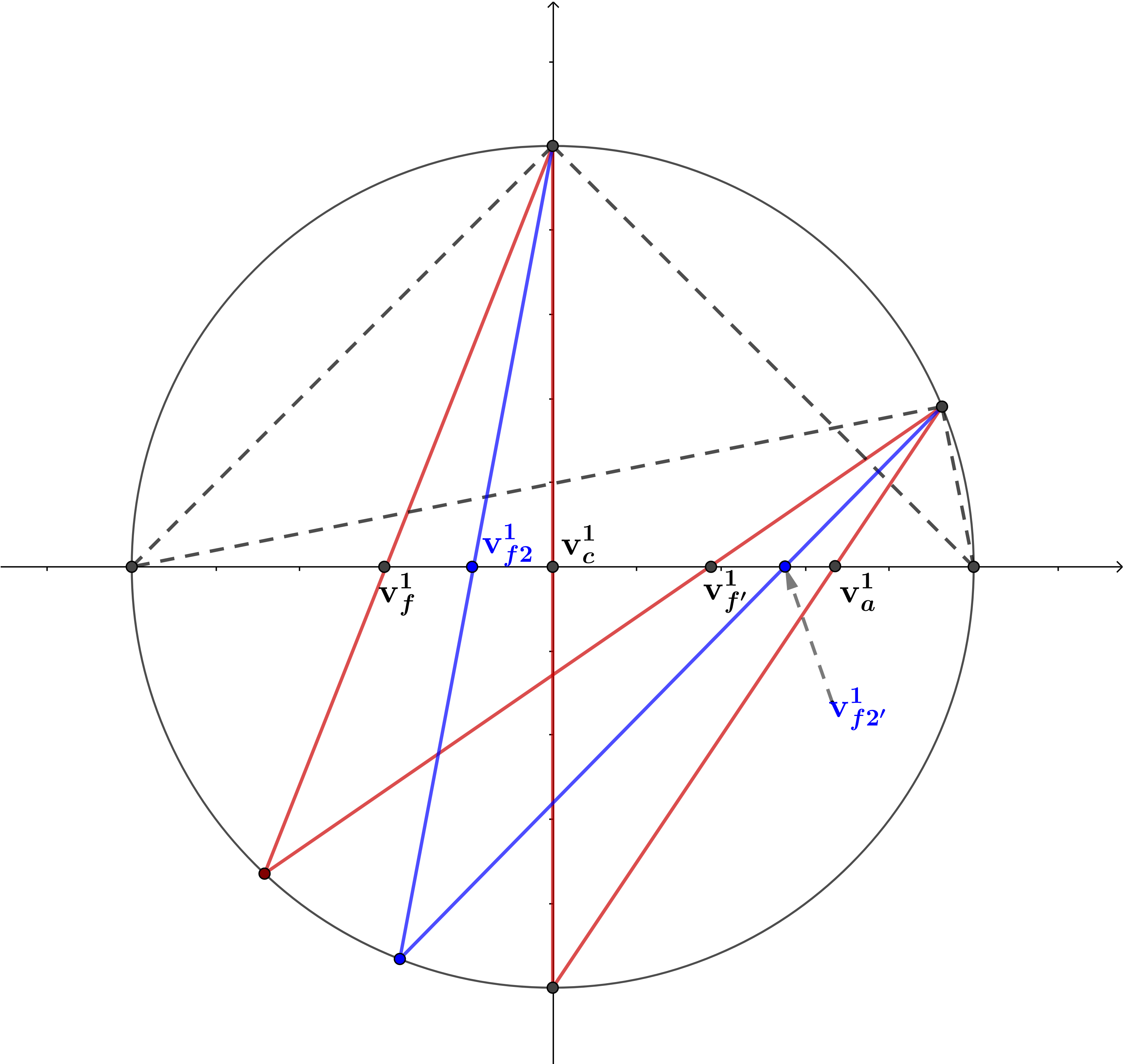}\label{CA_Cross2}}
	\caption{Invariance of the Hilbert distance under observer changes: illuminants $C$ and $A$, and colors $F$, $F'$, $F_2$, and $F_2'$.}
	\label{HilbertCA2}
\end{figure}

The same reasoning applied to the situation depicted in Fig. \ref{CA_CIE2}, where the points $F_2$ and $F_2'$ belong to another iso-chroma contour, leads to:
\begin{equation}
	d_H (\textbf{v}_{1F_2},\textbf{v}_{1c}) = d_H (\textbf{v}_{2F_2'},\textbf{v}_{2a})= d_H (\textbf{v}_{1F_2'},\textbf{v}_{1a})\ ,
	\label{CA_Cross_Chasles2}
\end{equation}
see Fig. \ref{CA_Cross2}.

\begin{figure}[h!]
	\centering 
	\subfigure[The three illuminants $C$, $A$ and $G$, and the colors $F$ and $F'$, $F_2$ and $F_2'$, and $H$ and $H''$ in the $xyY$ diagram.]{\includegraphics[width=0.47\linewidth]{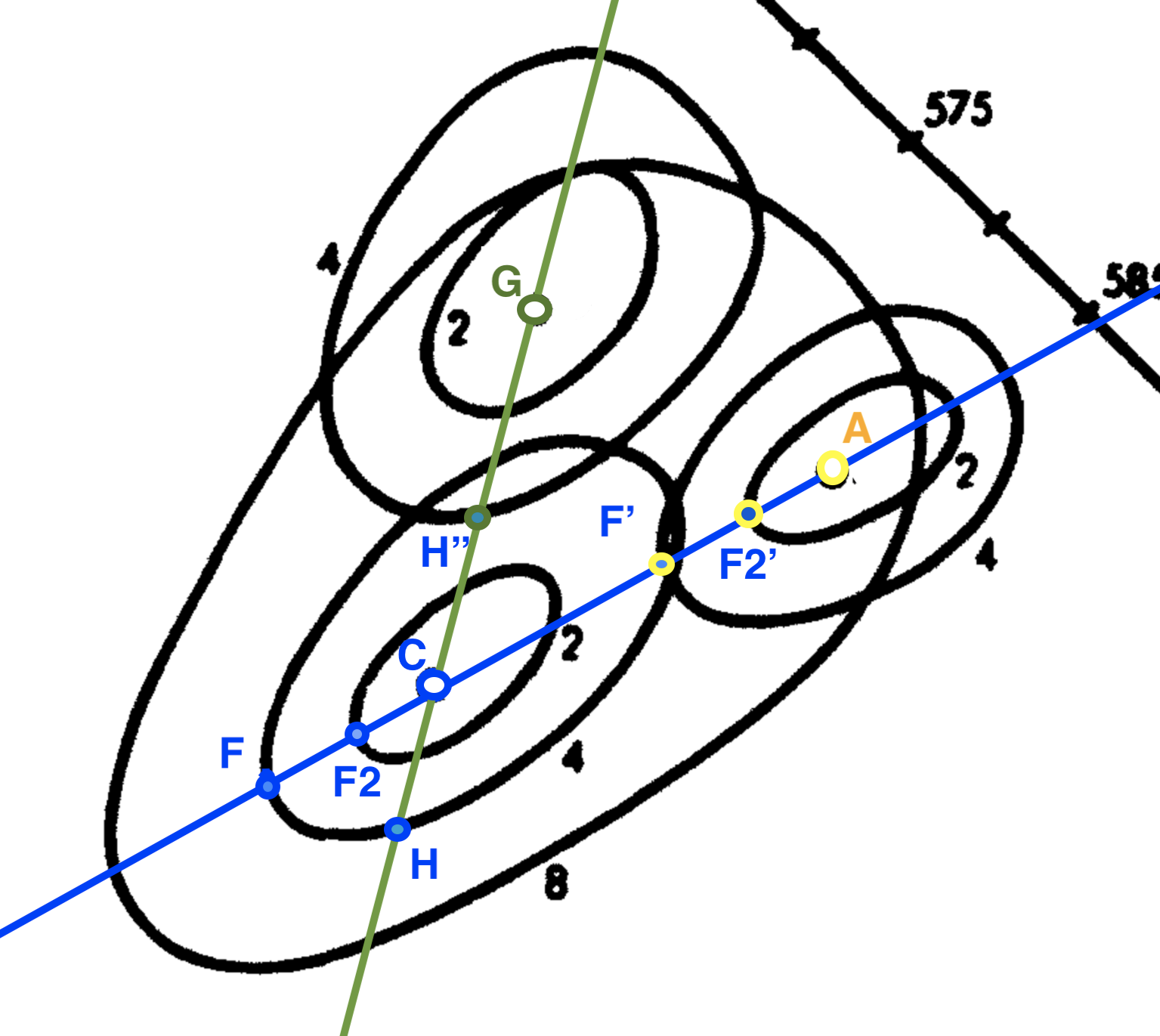}\label{CG_CIE}}
	\hspace{5mm}
	\subfigure[Illustration of the equalities of Eq. (\ref{CG_Cross_Chasles}) in the disk $\mathcal{D}_{1/2}$.]
	{\includegraphics[width=0.42\linewidth]{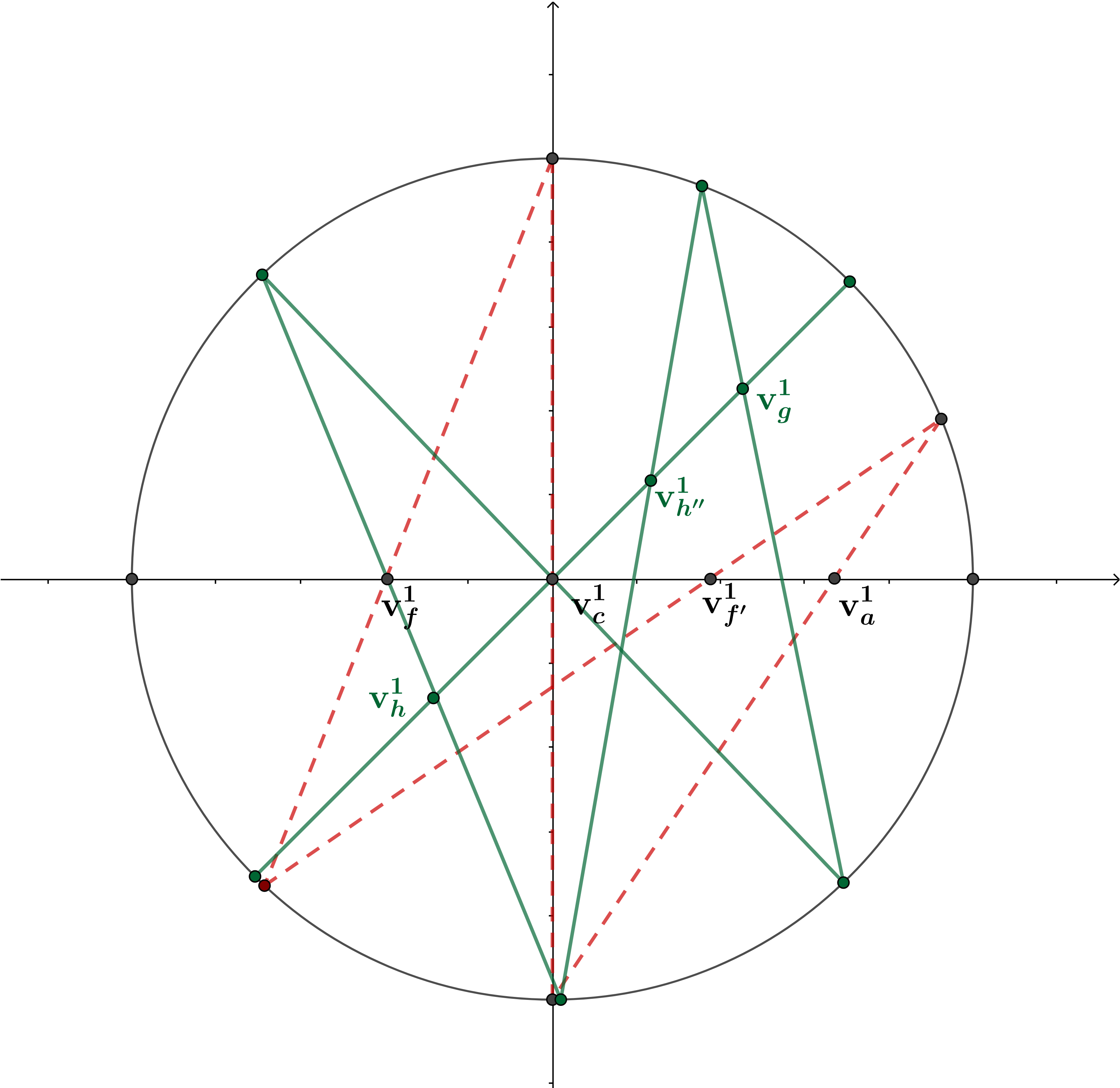}\label{CG_Cross}}
	\caption{Invariance of the Hilbert distance under observer changes: illuminants $C$ and $G$, and colors $H$ and $H''$, compared with illuminants $C$ and $A$, and colors $F$ and $F'$.}
	\label{HilbertCG}
\end{figure}

Finally, we consider the quite more complicated situation depicted in Fig. \ref{CG_CIE}. It is precised in \cite{Burnham:57} that `{\em A change from a blue ($C$) adaptation to a yellow ($A$) adaptation shows vectors running in a blue-yellow direction, a change from a blue ($C$) adaptation to a green ($G$) adaptation shows vectors running in a blue-green direction.}' This means that the angle between ${\bf v}_{1a}$ and ${\bf v}_{1g}$ is equal to $\pi/4$. From Fig. \ref{CG_CIE} we can approximate the norm of the chromatic vector ${\bf v}_{1g}$: $\Vert {\bf v}_{1g}\Vert\simeq 0.32.$ The chromatic vectors ${\bf v}_{1H}$ and ${\bf v}_{3H''}$ of the two colors $H$ and $H''$ marked on Fig. \ref{CG_CIE} are equal. Once again, one can easily check that:
\begin{equation}
	d_H ({\bf v}_{1H},{\bf v}_{1c}) = d_H ({\bf v}_{3H''}, {\bf v}_{3g})= d_H ({\bf v}_{1H''},{\bf v}_{1g})\ , 
	\label{CG_Cross_Chasles}
\end{equation}
see Fig. \ref{CG_Cross}.

These discussions show clearly that the Hilbert metric is compatible with the only psychovisual data that we have at disposal. Here we have reported only three cases, but other three configurations related to Fig. \ref{fig:IsoChroma} can be studied and our computations showed that they give rise to the same conclusions. We have only treated the case when colors, e.g. $F$ and $F'$, have chromatic vectors collinear to the new observer chromatic vector, e.g. ${\bf v}_{1F}$ and ${\bf v}_{1F'}$ are collinear to ${\bf v}_{1a}$ in this first situation. Dealing with arbitrary colors needs the introduction of more sophisticated mathematical tools to take into account the general addition law for non-collinear vectors. We prefer to not enter in such details in the present work and to postpone the general case for future research.

\section{Chromatic aberration, boost maps and the theoretical derivation of the outcome of Yilmaz's third experiment}\label{sec:boostaberration}
In this final section we show how eq. (\ref{third}), the third and final result quoted by Yilmaz, can be obtained as a theoretical consequence of the trichromacy axiom. We will obtain this result by explaining how to recover the chromatic aberration effect from it, which happens to be related to how pure chromatic states generate Lorentz boost maps that act on the space of chromatic vectors.     
\subsection{One parameter subgroups of boost maps} Let us recall that the pure chromatic states are given by density matrices of the form:
\begin{equation}
	\rho({\bf v})={1\over 2}(Id_2+{\bf v}\cdot\sigma)={1\over 2}\left(\begin{array}{cc}1+v_1 & v_2 \\v_2 & 1-v_1\end{array}\right)\ ,
\end{equation}
where ${\bf v}=(v_1,v_2)$ is a unit vector of $\mathbb R^2$. We have the following result, see also \cite{Berthier:2020}.

\begin{proposition} Every pure chromatic state generates a one-parameter subgroup of Lorentz boosts.
\end{proposition}

\medskip

\proof The matrix 
\begin{equation}
	A({\bf v},\zeta_0)=\exp\left(\zeta_0{{\bf v}\cdot \sigma\over 2}\right)\ ,
\end{equation}
with $\zeta_0$ a real parameter, is an element of the group $PSL(2,\mathbb R)$. Using the action of  $PSL(2,\mathbb R)$ on $\mathcal H(2,\mathbb R)$ we can consider the matrices given by:
\begin{equation}
	\sigma_i\longmapsto A({\bf v},\zeta_0)\sigma_i A({\bf v},\zeta_0)\ ,
\end{equation}
for $i=0,1,2$ with $\sigma_0=Id_2.$ The matrix with entries
\begin{equation}
	M({\bf v},\zeta_0)_{ij}={1\over 2}{\rm Trace}\left(\sigma_iA({\bf v},\zeta_0)\sigma_jA({\bf v},\zeta_0)\right)\ ,
\end{equation}
is the matrix
\begin{equation}
	\label{boost}
	M(\zeta)=\left(\begin{array}{ccc}\cosh(\zeta_0)&v_1\sinh(\zeta_0)  &  v_2\sinh(\zeta_0) \\ v_1\sinh(\zeta_0) & 1+v_1^2(\cosh(\zeta_0)-1) &  v_1v_2(\cosh(\zeta_0)-1)\\ v_2\sinh(\zeta_0)  &  v_1v_2(\cosh(\zeta_0)-1)& 1+v_2^2(\cosh(\zeta_0)-1) \end{array}\right)\ ,
\end{equation}
with $\zeta=\tanh(\zeta_0)(v_1,v_2).$
\qed

If, in particular, $\textbf{v}=(1,0)$, then 
\begin{equation}
	M(\zeta)=\left(\begin{array}{ccc}\cosh(\zeta_0)&\sinh(\zeta_0)  &  0 \\ \sinh(\zeta_0) & \cosh(\zeta_0) &  0\\ 0  &  0 & 1 \end{array}\right)\ .
	\label{exampleboost}
\end{equation}
One can easily check that, in this case, the pure pure chromatic vector $(\cos\theta,\sin\theta)/2$ is sent to the pure chromatic vector ${\bf w}=(w_1,w_2)$, where 
\begin{equation}
	\left\{
	\begin{array}{ll}
		2w_1=\displaystyle{\tanh(\zeta_0)+\cos\theta\over 1+\tanh(\zeta_0)\cos\theta}\\
		2w_2=\displaystyle{(1-\tanh(\zeta_0)^2)^{1/2}\sin\theta\over 1+\tanh(\zeta_0)\cos\theta}\ .
	\end{array}
	\right.
	\label{aberration}
\end{equation}

\subsection{A theoretical derivation of the outcome of Yilmaz's third experiment}

The following discussion show how eq. (\ref{aberration}) can be used to derive the outcome of the third Yilmaz experiment, eq. (\ref{third}). The pure chromatic vector $(0,1)/2$ is sent to the pure chromatic vector with coordinates $(\tanh(\zeta_0),(1-\tanh(\zeta_0)^2)^{1/2})/2$ whereas the pure chromatic vector $(1,0)/2$ remains unchanged. 
When the rapidity $\zeta_0$ increases, $\tanh(\zeta_0)$ approaches 1 and the vector $(\tanh(\zeta_0),(1-\tanh(\zeta_0)^2)^{1/2})/2$ approaches the vector $(1,0)/2$. At the limit $\tanh(\zeta_0)=1$, every pure chromatic vector $(\cos\theta,\sin\theta)/2$ is sent to the vector $(1,0)/2$, except the vector $(-1,0)/2$.

This means that every pure chromatic vector, except the green pure chromatic vector, can be transformed to a pure chromatic vector arbitrarily close to the red pure chromatic vector under the Lorentz boost of eq. (\ref{exampleboost}) if the rapidity $\zeta_0$ is sufficiently large. 

Equation (\ref{aberration}) allows us to provide a theoretical explanation of the results of Yilmaz's third experiment. To this aim, note that $w_1$ is the cosine of the angle of the ray from the achromatic vector to the image of the chromatic vector $(\cos\theta,\sin\theta)/2$ viewed under the initial illuminant $I$, whereas 
\begin{equation}
	\overline w_1={-\tanh(\zeta_0)+\cos\theta\over 1-\tanh(\zeta_0)\cos\theta}
\end{equation}
is the cosine of the angle of the ray from the achromatic vector to the image of the chromatic vector $(\cos\theta,\sin\theta)/2$ viewed under the illuminant $I'$. As a consequence, under the illuminant $I'$, the expected yellow chromatic vector given by $\theta=\pi/2$ is in fact the greenish chromatic vector given by $\cos\theta=-\tanh(\zeta_0)$.

\section{Discussion}\label{sec:discussion}
In this paper we have strengthen the novel quantum theory of color perception proposed in \cite{Berthier:2020} and we extend it to incorporate also relativistic phenomena, resulting in a coherent relativistic quantum theory of color perception.

We have shown that the noticeable, yet heuristic, intuition of Yilmaz \cite{Yilmaz:62} regarding the relativistic nature of color perception can be incorporated in a rigorous mathematical setting that can be built from the single axiom of trichromacy. 

We have obtained this result by following the hint given by Mermin's alternative, and perhaps more profound, reconstruction of the special theory of relativity from Einstein-Poincaré addition law for velocity vectors. This led us to define and analyze the crucial concept of perceptual chromatic vector and to show that such vectors actually satisfy Einstein-Poincaré addition law. 

Quite surprisingly, this fact also allowed us to coherently endow the space of perceptual chromatic vectors with the Hilbert metric, which we verify to be in accordance with known  experimental results, thus underline the importance of such a distance in color perception. 	

We consider fascinating that both the relativistic and the quantum components of the theory of color perception that we describe in this paper are based on unconventional and quite rarely used approaches: Mermin's viewpoint on special relativity and Jordan's algebraic perspective on quantum theories.

Besides these approaches, one can envisage to focus on a more emblematic aspect of quantum theory and try to describe color perception from quantum measurements. This can be done by considering the effect space of the rebit introduced in section \ref{sec:quantcolopp}. Here we give only some ideas of how to proceed since a more complete study will appear in a future paper.

In quantum information, the concept of an effect refers to a measurement apparatus that produces an outcome. The duality between states and effects means essentially that when a state and an effect are specified, one can compute a probability distribution which is the only meaningful information that we can obtain about the experiment. One can check that the effect space of the rebit is the set
\begin{equation}
	\mathcal{E(\mathcal D)}= \left\{ e \in  \mathcal {C^*(\mathcal D)} , e \leq Id\right\},
\end{equation}
where 
\begin{equation}
	\mathcal C(\mathcal D)=\{\alpha(1,{\bf v}), \alpha\geq 0, {\bf v}\in\mathcal D\}
\end{equation}
is the state cone of the unit disk $\mathcal D$ (see section \ref{sec:quantcolopp}). Remarkably, this effect space, when represented in the 3-dimensional Minkowski space, coincides precisely with the perceptual double cone depicted by de Valois and de Valois in \cite{Devalois:97}. 

Effects appear naturally when considering generalized measurements. More precisely, given a generalized measurement described by a set $\{M_m\}_m$ of $2\times2$ real matrices satisfying $\sum_mM_m^tM_m=Id$, if we denote $E_m=M_m^tM_m$, then $E_m\in\mathcal H^+(2,\mathbb R)$ is an effect.

Now, if $\rho$ is a state density matrix, the real number $p(m)={\rm Tr}(E_m\rho)$ is the probability of the outcome $m$ of the generalized measurement $\{E_m\}$ evaluated on the state given by $\rho$. The unrescaled post-measurement state is given by
\begin{equation}
	\rho_m=M_m\rho M_m^t.
\end{equation}
Note that 
\begin{equation}
	{\rm Tr}(\rho_m)=p(m)={\rm Tr}(E_m\rho)
\end{equation}
so that $\rho_m$ may be considered as a generalized state density matrix whose trace belongs to the interval $[0,1]$.

Using the isomorphism of Jordan algebras\footnote{$\sigma_0=Id_2$ and Einstein's summation convention is implicitly adopted.}
\begin{gather}
	\varphi:\mathcal H(2,\mathbb R)\longrightarrow \mathbb R\oplus\mathbb R^2\simeq \mathbb R^{1,2}\\
	\varpi= {1\over 2}\varpi^i\sigma_i\longmapsto {1\over 2}(\varpi^0, \varpi^1e_1+\varpi^2e_2)\simeq{1\over 2}\varpi^ie_i\equiv \underline\varpi\nonumber,
\end{gather}
and considering the map (spinor representation)
\begin{gather}
	\psi : \mathcal M(2,\mathbb R)\longrightarrow \mathcal L(\mathbb R^{1,2}, \mathbb R^{1,2})\\
	A\longmapsto \varphi\circ {\rm Ad}_A\circ\varphi^{-1}\nonumber
\end{gather}
such that 
\begin{equation}
	\psi(A)(\underline\varpi)=\underline{A\varpi A^t},
\end{equation}
one can verify that
\begin{equation}
	\underline{\rho_m}=\psi(U_m)\left(\underline{\sqrt{E_m}\rho\sqrt{E_m}}\right)=\psi(U_m)\circ\psi\left(\sqrt{E_m}\right)(\underline\rho),
\end{equation}
where $M_m=U_m\sqrt{E_m}$ with $U_m\in\text{O}(2).$

The transformation $\psi(U_m)$ is nothing but a 3-dimensional rotation. The transformation $\psi\left(\sqrt{E_m}\right)$ is given by 
\begin{equation}
	\psi\left(\sqrt{E_m}\right)=\Vert\underline E_m\Vert_{_\mathcal M}L(v)
\end{equation}
where $L(v)$ is the Lorentz boost parametrized by the chromatic vector $v=(v_1,v_2)$ of the effect $E_m$, and $\Vert\underline E_m\Vert_{_\mathcal M}$ is the Minkowski norm of $\underline{E}_m$.

These computations show that the image $\underline{\rho_m}$ of the post-measurement state $\rho_m$ in the 3-dimensional Minkowski space is the image under the Lorentz boost $L(v)$ of $\underline\rho$, provided that $U_m$ is the identity\footnote{Which means that there is no evolution due to the environment.}. As a consequence,  the relativistic interpretation that we have derived from our quantum model is also very coherent with quantum information processes.

We are currently investigating the mathematical details of the extension of our proposal from collinear chromatic vectors to the general case in which chromatic vectors do not necessarily lie on the same axis. Moreover, we are also analyzing the possibility to explain well-known perceptual effects, see e.g. \cite{Fairchild:13} and \cite{Gronchi:17}, with the theoretical framework discussed here. 

We also consider interesting to study how the novel objects and formalism that we have introduced in this work can be used for practical colorimetric purposes and how they relate to existing color spaces represented in cylindrical coordinates such as the HSV space. To this aim, and also to more finely test our proposal, it is paramount to complement the exiguous psychovisual data that we currently have at disposal and possibly to design new kind of experiments.

Finally, we deem that a significant part of our contribution is also represented by a  whole new  nomenclature that we introduced in sections \ref{sec:defquantperceptual}, \ref{sec:nomenclaturespecialcolor} and \ref{sec:colorimetricdefs}. This is an unavoidable step when building, or refining, a novel theory. Nevertheless, in doing so we have tried to remain as close as possible to terms already present in quantum mechanics, special relativity and standard colorimetry, respectively.

\section{Appendix - Description of Yilmaz experiments}\label{sec:appendix}
\begin{figure}[htbp]
	\centering
	\includegraphics[scale=0.13]{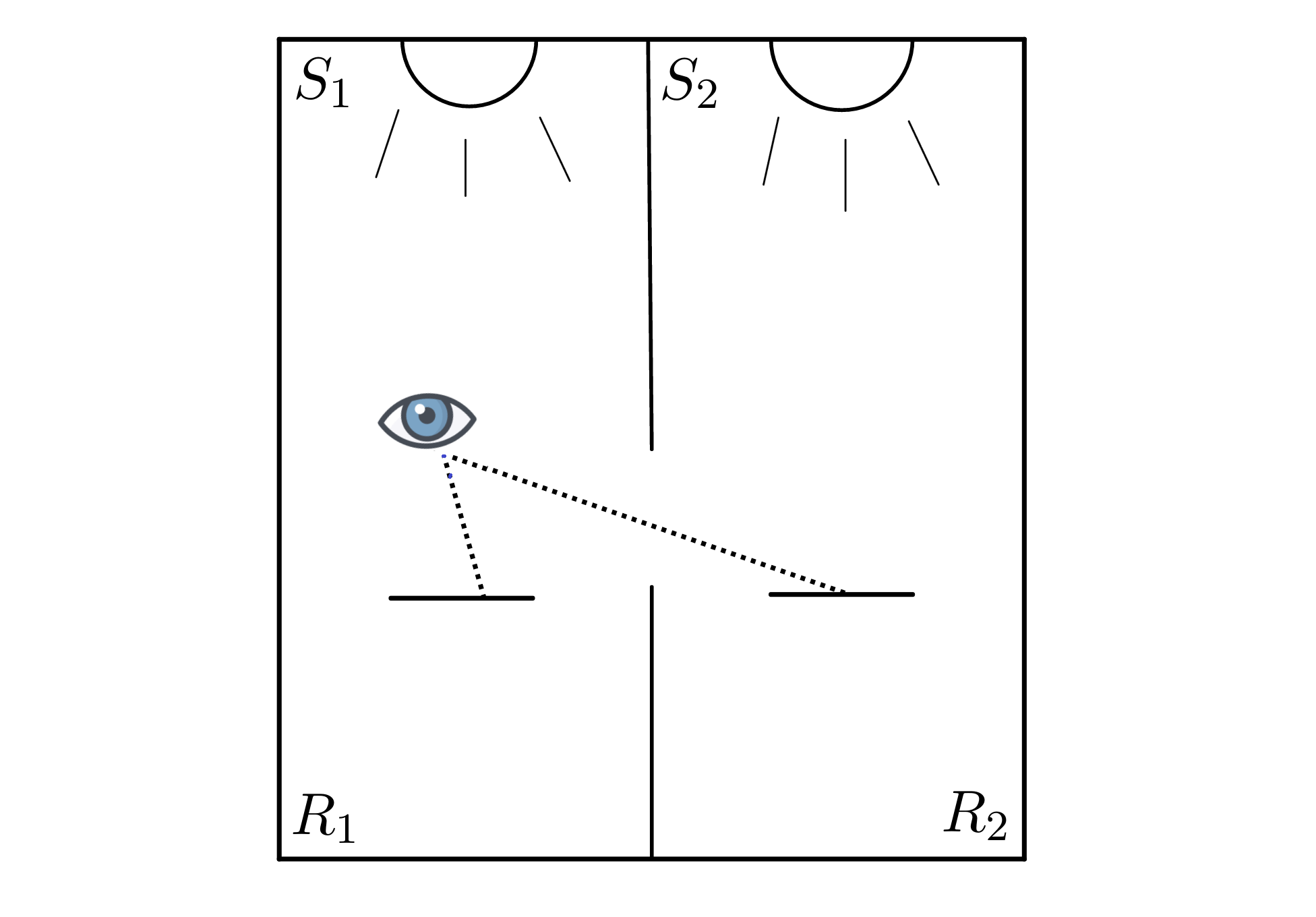}\ \ \includegraphics[scale=0.45]{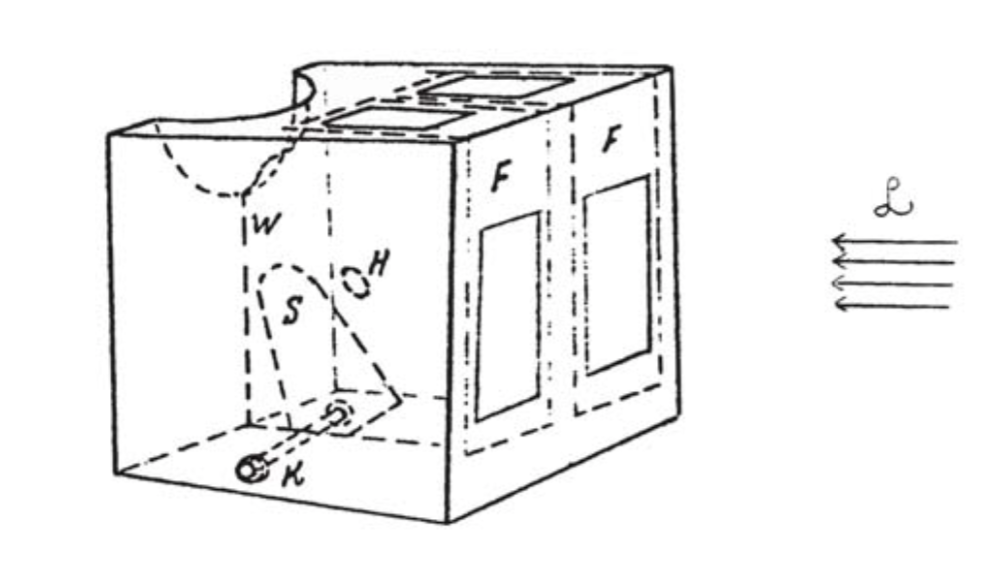}
	\caption{The experimental apparatus considered by Yilmaz. The image on the right is from Inter-Society Color Council News, Issue 419, Jan/Feb 2006, by kind concession of M. H. Brill, whom we would like to thank for sharing this reference with us.}\label{fig:figure7}
\end{figure}

The generic apparatus for the experiments is shown in Fig. \ref{fig:figure7}, where we can see two identical rooms $R_1$ and $R_2$, separated by a common wall with a thin hole and illuminated by the sources of light $S_1$ and $S_2$. Both rooms are painted with a non-selective Lambertian white paint. A piece of white paper is divided in two parts and they are placed in the rooms, so that an observer can perceive them simultaneously. The key point is that one piece is seen directly and the other through the hole.

The illumination $S_1$ of room $R_1$ will always be provided by near-daylight broadband illuminants.  Instead, the illumination of room $R_2$ will be provided by a light source $S_2$ that can also be narrow-band.
The perceived colors are compared with the help of a set of Munsell chips enlighted by the same illuminant under which the observer is adapted. 

A detailed analysis of the interpretation and feasibility of the experiments is available in \cite{Prencipe:20}, here we quote directly Yilmaz \cite{Yilmaz:62} to allow the reader to make up her or his own mind about them.

\subsection{The first experiment}
`{\em If the sources $S_1$ and $S_2$ are chosen to be two different illuminants of near-daylight chromaticity, $I$ and $I'$, then the wall of each room is perceived as white by the observer in the room but the wall of the other room, as seen through the hole, appears chromatically colored. Furthermore, if $R_2$ appears with the saturation $\sigma$ from $R_1$, then $R_1$ appears with the saturation approximately $-\sigma$ from $R_2$, the minus sign indicating that the hue is complementary to the former hue.}'

\subsection{The second experiment}
`{\em If $S_2$ is chosen to be a single-frequency source, say, corresponding to the long-wave (red) extreme of the visible spectrum $\overline R$, then the saturation $\Sigma$ observed through the hole (observer being in $R_1$) is too high to be duplicated by any of the Munsell chips, and remains practically the same if we change the illuminant from $I$ to $I'$ in $R_1$.}'

\subsection{The third experiment}
`{\em Let $S_2$ be a source of frequency corresponding to the yellow part of the spectrum, $\overline Y$, separated in the hue circle by 90 degrees from spectrum red, $\overline R$. Then if we change the illuminant in $R_1$ from $I$ to $I'$, the hue of $\overline Y$ is seen to change by an amount $\varphi$ such that $\sin\varphi\simeq \sigma/\Sigma$. No variation seems to take place in its saturation.}'

\section*{Acknowledgments}
This study has been carried out with financial support from the French State, managed by the French National Research Agency (ANR) in the frame of the “Investments for the future” Programme IdEx Bordeaux-SysNum (ANR-10-IDEX-03-02). We also acknowledge partial support by the Nouvelle Aquitaine Region (convention 2018-1R50104), by the French CNRS with the grant 80 primes and the grant GoalVision, and by Huawei Technologies France SASU.

The authors wish to thank the anonymous referees for their interesting suggestions that permitted us to improve this work, especially in what concerns the relevance of our quantum model regarding the state of the art on the use of hyperbolic metrics in color perception.

\bibliographystyle{plain} 
\bibliography{bibliography}

\end{document}